\documentclass[pra,longbibliography,floatfix,superscriptaddress,reprint,]{revtex4-1}
\usepackage{amsmath}
\usepackage{graphicx}
\usepackage{dcolumn}
\usepackage{xcolor}   
\usepackage{bm}        
\usepackage{amssymb}   
\usepackage{physics}
\usepackage{tikz}
\usepackage{gensymb}
\usepackage{xcolor}
\usepackage{float}
\usepackage{dsfont}
\usepackage[caption = false]{subfig}
\usepackage{mathtools}
 \usepackage{amsthm}
 \usepackage[english]{babel}
\usepackage[normalem]{ulem}
\usepackage{cancel}

\usepackage[bookmarks,bookmarksopen,bookmarksdepth=2]{hyperref}
\hypersetup{
    colorlinks=true,
    citecolor=blue,
    linkcolor=blue,
    filecolor=magenta,
    urlcolor=blue,
}
 \usepackage{quantikz}
\usepackage[capitalize]{cleveref}
\usepackage{babel}
\crefname{section}{Sec.}{Secs.}  
\crefname{appendix}{App.}{Apps.}  
\usepackage[english]{babel}

\newcommand{\stkout}[1]{\ifmmode\text{\sout{\ensuremath{#1}}}\else\sout{#1}\fi}

\makeatletter
\DeclareRobustCommand{\element}[1]{\@element#1\@nil}
\def\@element#1#2\@nil{%
  #1%
  \if\relax#2\relax\else\MakeLowercase{#2}\fi}
\makeatother
\let\oldaddcontentsline\addcontentsline

\newcommand{\starttocentries}{\let\addcontentsline\oldaddcontentsline}

\newcommand{\nocontentsline}[3]{}
\newcommand{\tocless}[2]{\bgroup\let\addcontentsline=\nocontentsline#1{#2}\egroup}

\begin{document}
\widetext

\title{Fault-tolerant quantum computation using large spin cat-codes}
\def \addCQuIC {Center for Quantum Information and Control, University of New Mexico, Albuquerque, NM, USA}
\def \addSandia {Sandia National Laboratories, Albuquerque, NM, USA}
\def \addPandAUNM {Department of Physics and Astronomy, University of New Mexico, Albuquerque, NM, USA}
\def \addPandAUO {Department of Physics and Astronomy, University of Oklahoma, Norman, OK, USA}
\def \addLANL {Los Alamos National Laboratory, Los Alamos, NM, USA}
\def \addECEUNM{Department of Electrical and Computer Engineering,
University of New Mexico, Albuquerque, New Mexico 87131, USA}
\author{Sivaprasad Omanakuttan}
\email[]{somanakuttan@unm.edu}
\affiliation{\addCQuIC} \affiliation{\addPandAUNM}
\author{Vikas Buchemmavari}
\affiliation{\addCQuIC} \affiliation{\addPandAUNM}
\author{Jonathan A. Gross}
\affiliation{Google Quantum AI, Venice, CA 90291, USA
}
\author{Ivan H Deutsch}
\email[]{ideutsch@unm.edu}
\affiliation{\addCQuIC} \affiliation{\addPandAUNM}
\author{Milad Marvian}
\email[]{mmarvian@unm.edu}
\affiliation{\addCQuIC} \affiliation{\addPandAUNM} \affiliation{\addECEUNM}

\date{\today}
\begin{abstract}
We construct a fault-tolerant quantum error-correcting protocol based on a qubit encoded in a large spin qudit using a spin-cat code, analogous to the continuous variable cat encoding.
With this, we can correct the dominant error sources, namely processes that can be expressed as error operators that are linear or quadratic in the components of angular momentum.  
Such codes tailored to dominant error sources {can} exhibit superior thresholds and lower resource overheads when compared to those designed for unstructured noise models. 
A key component is the CNOT gate that preserves the rank of spherical tensor operators. 
Categorizing the dominant errors as phase and amplitude errors, we demonstrate how phase errors, analogous to phase-flip errors for qubits, can be effectively corrected. Furthermore, we propose a  measurement-free error correction scheme to address amplitude errors without relying on syndrome measurements.
Through an in-depth analysis of logical CNOT gate errors, we establish that the fault-tolerant threshold for error correction in the spin-cat encoding surpasses that of standard qubit-based encodings.
We consider a specific implementation based on neutral-atom quantum computing, with qudits encoded in the nuclear spin of $^{87}$Sr, and show how to generate the universal gate set, including the rank-preserving CNOT gate, using quantum control and the Rydberg blockade. These findings pave the way for encoding a qubit in a large spin with the potential to achieve fault tolerance, high threshold, and reduced resource overhead in quantum information processing.
 \end{abstract}
\maketitle

\tocless{\section{Introduction}}{\label{sec:introduction}}


Quantum computers have the potential to provide a substantial advantage over their classical counterparts \cite{shor1999polynomial,farhi2001quantum,lloyd1996universal,biamonte2017quantum,aspuru2005simulated}. However, quantum computers are extremely susceptible to environmental noise and imprecise control, which hinders achieving their full computational capacity.
Fault-tolerant quantum computation (FTQC), provides a solution to perform reliable computation even in the presence of imperfect elementary components \cite{knill1998resilient,aharonov1997fault,knill2005quantum,raussendorf2007topological}. 
The cornerstone of FTQC is the threshold theorem, which states that if the error rate of individual components remains below a constant threshold, then arbitrarily long quantum computation can be performed \cite{aharonov1997fault,knill1998resilient,preskill1998reliable,kitaev1997quantum,Aliferis2008fault}. In addition to the value of noise threshold, a critical aspect of FTQC is the resource overhead, quantifying the number of physical systems required to encode logical information.
Despite the formidable challenges, there has been notable experimental progress in FTQC, bringing us closer to harnessing the full potential of quantum computing \cite{acharya2022suppressing,ryan2022implementing,krinner2022realizing,postler2022demonstration,Bluvstein_Lukin_2023_QEC_Logical,gupta2024encoding}. 

The conventional approaches for FTQC are mostly devoted to structureless and uncorrelated noise. 
An instance of this is depolarizing noise, where all local Pauli operators have an equal probability. 
However, such decoherence models often entail stringent threshold requirements and result in significant overheads for FTQC
~\cite{knill2005quantum,raussendorf2007topological,svore2006noise,spedalieri2008latency}.
An alternative strategy involves seeking error-correcting codes tailored to the prevalent noise sources of the particular physical platform.
When possible, these tailored approaches can lead to improved thresholds and reduced resource overhead~\cite{Aliferis2008fault,zzPoulin}. 
For instance, biased qubits in bosonic systems can lead to exponentially suppressed bit flip errors compared to phase flip error~\cite{Cong_Lukin_QEC_Rydberg_PRX_2022,ofek2016extending,puri2020bias,guillaud2019repetition}. Additionally, in scenarios where erasure errors dominate over Pauli errors,  tailored error-correcting codes have proven advantageous~\cite{Grassl_erasure_1997_PRA,wu2022erasure,Sahay_Puri_biased_erasure_PRX_2023}
. By addressing the specific characteristics of dominant noise sources, these tailored methods offer promising avenues to enhance the performance of FTQC.

Another avenue to develop more efficient error-corrected quantum processors is to take advantage of the larger Hilbert spaces that can be controlled in individual subsystems for a given physical platform.
While many platforms offer access to multiple levels, the focus is often on isolating two well-defined levels for qubit-based computations. However, a more advantageous approach emerges when we exploit these multiple levels to create qubits naturally resilient to dominant noise channels \cite{GKP_Gottesman_PRA_2001,gross2021hardware,omanakuttan2023multispin,PhysRevA.108.022428,PhysRevA.85.040306}.  
The quintessential example is the Gottesman-Preskill-Kitaev encoding of a qubit in an infinite dimensional oscillator \cite{GKP_Gottesman_PRA_2001}. 
In this work we will consider encoding a qubit in a spin-$J$ system, corresponding to a qudit with $d=2J+1$ levels \cite{omanakuttan2021quantum,omanakuttan2022qudit,zache2023fermion}. 
By harnessing the properties of this qudit with multiple levels, we can establish logical qubits that possess inherent resistance to the impact of dominant noise channels, paving the way for more robust quantum computation.


Other works have previously explored the concept of encoding a qubit in a large spin~\cite{Gross2021, gross2021hardware, omanakuttan2023multispin}.
In this context, the angular momentum operators form the natural set of error operators for such encodings, generalizing the Pauli operator basis for qubits. Earlier studies identified quantum error-correcting encodings, but these constructions were not fault-tolerant~\cite{Gross2021,omanakuttan2023multispin}.  Here, our main objective is to investigate how we can achieve Fault-Tolerant Quantum Computation (FTQC), specifically for a qubit encoded in a large spin. 
This approach may be extended to a wide range of physical systems, including semiconductor systems \cite{Gross2021, gross2021hardware}, ion traps \cite{ringbauer2021universal, low2020practical}, atomic systems \cite{omanakuttan2021quantum, omanakuttan2022qudit, zache2023fermion}, molecules \cite{castro2021optimal,jain2023ae}, and superconducting systems \cite{ozguler2022numerical, blok2021quantum}, wherein spin qudits offer the means to encode logical qubits.

In this work, we direct our attention to a specific encoding we call the ``spin-cat encoding." 
This choice is motivated by the cat encodings employed in bosonic continuous variable systems~\cite{puri2020bias,guillaud2019repetition}, used to correct photon loss errors, the dominant errors for the continuous variable systems.  Similarly, spin-cat encoding can rectify the dominant error operators in spin systems, namely, the linear and quadratic angular momentum operators. Physically, these arise from uncontrolled Larmor precession of the spins and optical pumping between magnetic sublevels.
To achieve fault tolerance with spin-cat encoding, we develop two key ingredients.
First, we show how to implement a universal gate set that preserves the limited error space of interest.  An essential element here is the ``rank-preserving CNOT" gate that ensures that one does not convert correctable errors into uncorrectable ones. 
Second, aiming at a more easily implemented scheme, we develop a measurement-free error correction {gadget} for spin systems that require fresh ancilla spins {and data-ancilla operations but no measurements}. As we will show, this scheme effectively utilizes the rank-preserving CNOT gate in conjunction with standard phase flip error correction to address and correct angular momentum errors.

A distinctive aspect of the spin-cat encoding, setting it apart from other spin encodings~\cite{Gross2021,omanakuttan2023multispin,kubischta2023family,kubischta2023not}, is its unique structural composition.
In contrast to these earlier methods, the error subspaces in the spin-cat encoding partition the physical space into two-dimensional subspaces where logical operations act identically.
This gives the structure of a stabilizer code, a feature that plays a pivotal role in enabling fault-tolerant schemes for error correction.

The remainder of this article is organized as follows.
In \cref{sec:spin_cat_codes} we define the cat codes for spin systems and the natural basis for the dominant error channels. 
In \cref{sec:universal_gate_set}, {we discuss the requirements on gates in order to not spread correctable errors}. We describe the implementation of a rank-preserving CNOT gate for the encoding of a qubit in the nuclear spin of $^{87}$Sr in \cref{sec:CNOT_gate} and the necessary measurement and state preparation steps to implement the universal gate set in \cref{sec:state_preperation_and_measurement}.
In \cref{sec:syndrome_extraction}, we explain the protocol for syndrome extraction needed to correct the errors in spin-cat encoding, and the measurement-free error correction native to the qubit encoded in the spin. 
In \cref{sec:Fault_tolerance_and_threshold} {we obtain the threshold for FTQC based on the logical CNOT gate}. We conclude and explore possible future directions in \cref{sec:discussions_and_future_work}.
\vspace{0.7cm}

\tocless{\section{Generalization of cat code for Qudits/spin systems}}{
\label{sec:spin_cat_codes}}

\begin{figure*}
    \subfloat[]{\includegraphics[width =0.7\columnwidth]{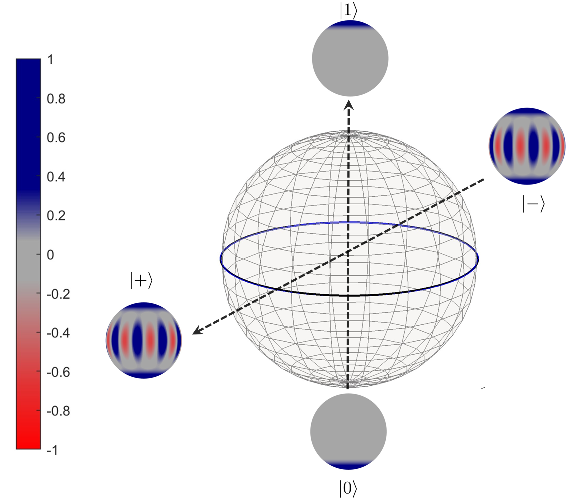} \label{fig:Fig_1_a}}
    \subfloat[]{\includegraphics[width=1.25\columnwidth]{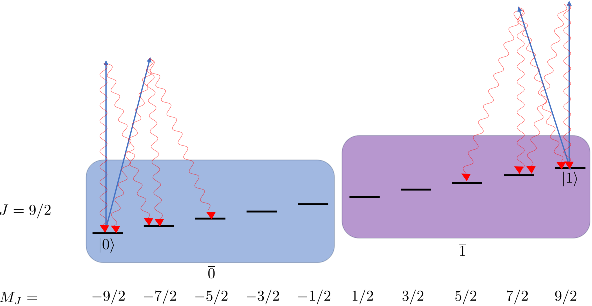}  \label{fig:Fig_1_b}}
    \caption{Qubit encoded in a spin using spin-cat states.
    (a) The Bloch sphere for the qubit encoded in a spin. The two spin-coherent states (stretched states) are the computational basis states, lying on the $Z$-axis and the spin-cat states then lie along the $X$-axis.  The spin Wigner function of the states is shown and its strong negativity indicates that spin-cats are highly nonclassical. (b) The spin-cat encoding of a qubit in spin $J=9/2$, $d=2J+1=10$ levels.
    The correctable errors divide the qudit into two subspaces, $\bar{0}$ and $\bar{1}$, shown as blue and purple boxes, respectively.  One physical error channel is optical pumping, corresponding to the absorption of photons (blur arrows) followed by spontaneous emission (wavy red arrows), which can lead to amplitude damping.}
       \label{fig:bloch_sphere}
\end{figure*}

In this section, we introduce our encoding, present the most prevalent types of noises in spin systems, and look at how they affect an encoded qubit.
We consider quantum information encoded in large spins with angular momentum $J$, a qudit of dimension $d=2J+1$. The space of local errors on a spin system is spanned by the irreducible spherical tensor operators ${T}^{(k)}_q(J)$~\cite{varshalovich1988quantum}  which are {orthogonal} polynomials in the spin angular momentum components, $\{{J}_x, {J}_y, {J}_z \}$ of order $k$, with {$q$ ranging from $-k$ to $k$}. 
The qudit operator space is spanned by the basis of tensors from $k=1$ to $k=2J$. 
In most platforms, physical errors are associated with low rank-$k$ tensors for $J \gg 1$.
For example, erroneous Larmor precession caused by noisy magnetic fields are generated by the SU(2) algebra, or rank-1 tensors.  
When controlled by laser light, as in atomic systems, optical pumping arising from photon scattering can lead to rank-2 errors. 
Higher rank errors are rare, as they involve multi-photon processes or higher rank tensor perturbations. 
We thus design codes that can correct any errors in the space spanned by the Kraus operators in the set {of linear and quadratic spin operators} $\{{T}^{(1)}_q(J), {T}^{(2)}_q(J)\}$ \cite{omanakuttan2023multispin}. 
For $J\gg1${, this} is a substantially reduced error space (dimension 8) compared to the total space of all possible errors (dimension $(2J+1)^2-1$).

To design a spin-encoding that can efficiently correct this biased noise structure, we consider the bosonic cat encoding of a qubit~\cite{puri2020bias}. In this encoding, the  qubit states $\ket{0}$ and $\ket{1}$ are chosen to be, 
\begin{equation}
    \ket{C_{\alpha}^{\pm}}\propto \ket{\alpha}\pm \ket{-\alpha},
\end{equation}
where $\ket{\alpha}$ is a coherent state of a single bosonic mode, for e.g., a mode of a microwave cavity as in superconducting systems. 
When the dominant source of noise is photon loss, this encoding exhibits a biased noise channel where increasing the amplitude $\alpha$, exponentially suppresses bit flip errors when compared to phase flip errors. 
It has been shown that by using simple codes such as a repetition code to correct phase flips, one can take advantage of this bias in the noise to achieve significant improvement in the threshold for FTQC \cite{Aliferis2008fault,puri2020bias} for cat qubits. 

 In this work, we pursue a similar approach for finite-dimensional spin systems and consider the spin-cat encoding with,

\begin{equation}
    \begin{aligned}
  \ket{\pm}&\equiv \frac{1}{\sqrt{2}}\left(\ket{J,-J}\pm\ket{J,+J}\right),
    \end{aligned}
 \label{eq:spin_cats}
    \end{equation}
{where now $\ket{0}=\ket{J,-J}$ and $\ket{1}=\ket{J,J}$ are the spin coherent states along the physical quantization ($z$) axis. 
We call this the spin-cat encoding}. 
  Similar to previous works based on continuous variable bosonic cat states \cite{puri2020bias,guillaud2019repetition}, {the spin cat states are defined along the $1$-axis of the qubit Bloch sphere}; see \cref{fig:Fig_1_a}. Note that, unlike the coherent states in the continuous variable setting, the spin coherent states are perfectly orthogonal to each other.

Despite utilizing a similar encoding, there are significant differences between the dominant sources of noise and the easy-to-implement operations in the spin system compared to bosonic cats. Thus, this encoding requires the development of new error-correction procedures that we address in this work.  Central to the continuous variable cat encoding, as explored in~\cite{puri2020bias,guillaud2019repetition}, is the reduction in bit-flip errors. 
The key to this bias is the presence of an energy gap between the excited state manifold and the logical subspace, that scales with $\lvert\alpha\rvert^2$. While this encoding offers significant advantages compared to standard qubit-based encoding, the leakage to these excited states can have detrimental effects on the energy-protected qubits. Dissipative stabilization can be employed to overcome these errors \cite{PhysRevLett.128.110502}.  

In contrast, in spin-cat encoding we use an alternative approach for fault tolerance.
We consider a primary layer of encoding where we correct for the physically relevant errors and then use a second layer of concatenation to achieve fault-tolerant quantum computation.
We can achieve this because the physically relevant errors are a small subset of all the possible errors for the encoded qubit.  For the spin-cat encoding, these physically relevant errors are composed of spherical tensors of rank-1 and rank-2, as described above. 
The key goal of the first layer of the encoding is to correct for these rank-1 and rank-2 errors.  Our protocol is fault-tolerant because the universal gates and error correction performed in the first layer of encoding do not convert lower-rank spherical tensor operators to higher-rank operators.  We call this ``rank-preserving'' error correction.  It is a generalization of the bias-preserving error correction where the dominant error for the encoded qubit is a single Pauli-error. In the second layer of encoding, the relevant errors are Pauli errors on the logical qubit, which can be corrected by any standard error correction protocol.

\vspace{0.7cm}
\tocless{\subsection{Error characterization}
\label{subsec:Error_characterization}}
To categorize the  relevant errors that can be corrected for the spin-cat encoding, it is useful to define the generalized ``kitten states''  as,
\begin{equation}
    \begin{aligned}
    \ket{\pm}_m&=\frac{1}{\sqrt{2}}\left(\ket{0}_m\pm\ket{1}_m\right).
    \end{aligned}
\end{equation}
where,
\begin{equation}
\begin{aligned}
\ket{0}_m&=\ket{J,-J+m}\equiv\ket{-J+m},\\
\ket{1}_m&=\ket{J,J-m}\equiv \ket{J-m}.
\end{aligned}
\label{eq:spin_cat_codes}
\end{equation}
The case $m=0$ is the spin-cat state. 
The total Hilbert space of the spin-cat encoding decomposes to $d/2$ qubit subspace where each of the qubit subspaces is spanned by the kitten states $\ket{\pm}_m$.
 {(For the remainder of this article, we consider  $J$ to be half-integer, i.e., even $d$. 
The proposed schemes can be easily adapted for odd $d$ with minor modifications.)}.
Thus we can write,
\begin{equation}
\mathcal{H}_d=\bigoplus_{i=0}^{\frac{d}{2}}\mathcal{H}_2^{(i)},
\label{eq:decompostion_space}
\end{equation}
where each $\mathcal{H}_2$ is a kitten subspace and $\mathcal{H}_d$ is the total Hilbert space of the qudit.
These subspaces are preserved by rotations about the spin quantization $z$-axis and by $\pi$ pulses around axes in the equatorial plane that exchange $\ket{\pm J}$.

We also define the following  projectors onto $\bar{0}$ and $\bar{1}$ subspaces that define correctable errors,
\begin{equation}
\begin{aligned}
    \Pi_{\overline{0}}&= \sum_{k=0}^{\lfloor \frac{2J-1}{2}\rfloor}\ketbra{-J+k}{-J+k},\\
    \Pi_{\overline{1}}&= \sum_{k=0}^{\lfloor \frac{2J-1}{2}\rfloor}\ketbra{J-k}{J-k}.\\
    \end{aligned}
    \label{eq:projectors}
\end{equation}
See \cref{fig:bloch_sphere} for an illustration. 

The relevant errors on the spin-cat encoding that we aim to correct are a combination of amplitude and phase errors. 
The amplitude errors are defined by the following transformation,
\begin{equation}
\ket{\pm}_m \to \sum_{k=0}^{\lfloor \frac{2J-1}{2}\rfloor} c_k \ket{\pm}_k,
\label{eq:amplitude_error}
\end{equation}
where $c_k$ is an arbitrary complex number. 
The  phase error is given by the transformation,
\begin{equation}
\ket{\pm}_k \to \ket{\mp}_k.
\label{eq:phase_error}
\end{equation}
{Physically, these occur as follows.}  First, consider spin rotations,
\begin{equation}
\begin{aligned}
U_{Z}&=\exp(-i\theta J_z),\\
U_{X}&=\exp(-i\theta J_x).
\end{aligned}
\end{equation}
For $\theta \ll 1$ their actions action on the spin-cat states is
\begin{equation}
\begin{aligned}
U_{Z}\ket{\pm}&\approx \left(\mathds{1}-i\theta J_z\right) \ket{\pm} = \ket{\pm}-i\theta J \ket{\mp},\\
U_{X}\ket{\pm}&\approx \left(\mathds{1}-i\theta J_x\right) \ket{\pm} \\
&= \ket{\pm}-i\theta \frac{\sqrt{J}}{\sqrt{2}} \ket{\pm}_1.
\end{aligned}
\label{eq:error_prob_rotation}
\end{equation}
Thus, the effect of $U_{Z}$ is to introduce a phase error on the spin-cat states whereas $U_{X}$ generates an amplitude error that takes a cat state to a kitten state with $m=1$. 
The ratio of probabilities of amplitude errors to phase errors {due to random rotation errors} goes as $1/J$, and hence approaches zero for large values of $J$.

Next, we consider errors resulting from optical pumping associated with photon scattering. For example, given a laser photon linearly polarized along the quantization axis, followed by the emission of $q=0,\pm1$ helicity photon, the Lindblad jump (Kraus) operators $W_q$ are given by~\cite{Deutsch2000},  
\begin{equation}
    \begin{aligned}
        W_{0}&=\beta T^{(2)}_0,\\
        W_{+1}&=  i \alpha T^{(1)}_{-1}- \beta \sqrt{\frac{3}{4}}T^{(2)}_{-1},\\
          W_{-1}&= i \alpha T^{(1)}_{1}+\beta \sqrt{\frac{3}{4}}T^{(2)}_{1}.
    \end{aligned}
    \label{eq:jump_opt_simplified}
\end{equation}
where $\alpha, \beta$ are real numbers that depend on the atomic structure and the states being excited by a near resonance laser. (See \cref{sec:Ratio_optical_pumping_errors} details.) Optical pumping can include rank-2 tensors as it involves two photons. 
The effect of optical pumping introduces both amplitude errors that change the kitten subspace~\cref{eq:amplitude_error}, and phase errors as given in~\cref{eq:phase_error}.
In contrast to errors that result from rank-1 SU(2) rotation, in optical pumping, it is equally important to correct both amplitude damping and phase errors and ultimately, we must do so fault-tolerantly.

Amplitude errors up to rank $K=\lfloor 2J-1/2 \rfloor$ can be corrected by identifying whether the system is  in 
a specific kitten state with a given $m$ value.
To correct for the phase errors, we concatenate the spin-cat code {in} a repetition code  {with logical states},
\begin{equation}
\begin{aligned}
\ket{+_\mathrm{L}}&=\ket{+}\ket{+} \ket{+},\\
\ket{-_\mathrm{L}}&=\ket{-} \ket{-} \ket{-}.
\label{eq:concat_spin_cat}
\end{aligned}
\end{equation}
While we consider a three-qubit repetition code here and throughout \cref{sec:syndrome_extraction} for simplicity, in \cref{sec:Fault_tolerance_and_threshold} we will look at repetition codes with more than three qubits in order to calculate the threshold for fault-tolerance. One can then perform the corresponding error correction steps similar to the approach taken in the continuous variable encoding~\cite{puri2020bias,guillaud2019repetition}.   We call this the ``logical-level encoding" to differentiate it from the physical-level encoding in \cref{eq:spin_cat_codes}.  

More formally, in \cref{sec:KL_conditions} we show that the  logical-level encodings in \cref{eq:concat_spin_cat} can correct {any single spin} angular momentum errors of the form,
\begin{equation}
\begin{small}
\mathcal{E}_{K}=\left\{
J_x^{l}J_y^{m}J_z^{n}; 0\leq l+m+n\leq K=\lfloor\frac{2J-1}{2} \rfloor\right\}.
\label{eq:error_channel}
\end{small}
\end{equation}
In practice we can restrict our attention to quadratic polynomials.

\tocless{\subsection{The irreducible spherical tensor basis}}{\label{subsec:Spherical_tensor_basis}}

The irreducible spherical tensor basis provides a natural basis to characterize the action of the error operators. In the basis of the magnetic sublevels, the normalized tensors are~\cite{varshalovich1988quantum} 

\begin{equation}
    T^{(k)}_q(J) = \sqrt{\frac{2k+1}{2J+1}} \sum\limits_{m,m'=-J}^{J} C_{Jm';kq}^{Jm} \ketbra{J,m}{J,m'},\ 
\end{equation}
where $C_{J,m';kq}^{J,m}= \braket{J,m}{J,m';k,q}$ are the Clebsch-Gordan coefficients.
The spherical tensor operators of rank-$k$ are the solid harmonics consisting of polynomials on the angular momentum operators of order $k$.
To track how errors occur, it is convenient to introduce the following linear combination of the spherical tensor operators, 

\begin{equation}
    \begin{aligned}
        S^{(k)}_q(J)& = \frac{1}{\sqrt{2}}\left[T^{(k)}_q(J) + (-1)^k T^{(k)}_{-q}(J) \right],\\  
A^{(k)}_q(J) &=\frac{1}{\sqrt{2}}\left[T^{(k)}_q(J) - (-1)^k T^{(k)}_{-q}(J) \right],\\
S^{(k)}_0(J) &= T^{(k)}_0(J).
\label{eq:basis}
    \end{aligned}
\end{equation}
for  $0\leq k \leq 2J+1$  and $q>0$.
It is straightforward to check that these operators form another orthonormal basis for a spin-$J$ system, i.e.,

\begin{equation}
\begin{aligned}
   \Tr{\left(S^{(k)}_q\right)^{\dagger} S^{(k')}_{q'}}&=\Tr{\left(A^{(k)}_q \right)^{\dagger}{A}^{(k')}_{q'}}=\delta_{k,k'}\delta_{q,q'}, 
   \\ \Tr{\left(S^{(k)}_q\right)^{\dagger} A^{(k')}_{q'}}&=0,
\end{aligned}
\end{equation}
for $0\leq k,k'\leq 2J+1$, $0\leq q \leq k$, and $0\leq q' \leq k'$.
The action of the operators on the cat and kitten states are given ({for $q>0$}) as,

\begin{widetext}
    \begin{equation}
    \begin{aligned}
        S^{(k)}_q\ket{\pm}_l & =  \sqrt{\frac{2k+1}{2(2J+1)}} \left[(-1)^k C^{J,-J+l-q}_{J,-J+l;k,-q}\ket{\pm}_{l-q} +C^{J,-J+l+q}_{J,-J+l;k,q}\ket{\pm}_{l+q}\right],\\
         A^{(k)}_q\ket{\pm}_l&= \sqrt{\frac{2k+1}{2(2J+1)}} \left[(-1)^k C^{J,-J+l-q}_{J,-J+l;k,-q}\ket{\mp}_{l-q} -C^{J,-J+l+q}_{J,-J+l;k,q}\ket{\mp}_{l+q}\right],\\
         S^{(k)}_0\ket{\pm}_l&=
         \begin{cases}
         \sqrt{ \frac{2k+1}{2J+1}}  C^{J,-J+l}_{J,-J+l;k,0}\ket{\pm}_l, & \text{if }\ k \text{ mod } 2=0 \\
      \sqrt{ \frac{2k+1}{2J+1}}  C^{J,-J+l}_{J,-J+l;k,0}\ket{\mp}_l, & \text{otherwise.}
    \end{cases}
     \end{aligned}
     \label{eq:action_F_LM}
\end{equation}
\end{widetext}

Note that the states {on the righthand side of the equations} are not normalized, as the operators $S^{(k)}_q,A^{(k)}_q$ are not unitary.
They are the Kraus operators corresponding to the {relevant} errors.

The action of the Kraus operator $S^{(k)}_q$  is the amplitude error given in \cref{eq:amplitude_error}.
The Kraus operator $S^{(k)}_0$ flips the kitten states for $k\;\mathrm{mod}2$=1 which corresponds to the phase error in \cref{eq:phase_error}; the Kraus operators $A^{(k)}_q$ change the value of the kitten state and also flip their sign. 
This corresponds to the action of both amplitude and phase error.  This basis of the Kraus operators tracks whether the error is amplitude, phase, or the product of two.
The single spin errors that a spin-cat code can correct can be written in terms of the new basis as,
\begin{equation}
    \mathcal{E}_{K}=\left\{S_{q}^{(k)},{A}_q^{(k)} \; \vert \; 0\leq k\leq K ,-k\leq q\leq k\right\},
\label{eq:error_channel_basis}
\end{equation}
where $K=\lfloor\frac{2J-1}{2} \rfloor$.

The logical encoding defined in \cref{eq:concat_spin_cat} introduces a biased logical qubit so that the rate of bit flip errors is exponentially suppressed compared to the phase flip errors as a function of the total value of spin $J$. 
Any uncorrectable amplitude error at the physical level of the spin-cat encoding is transformed into a bit-flip error on the logical qubit. 
In \cref{fig:rotation_error_comparison} we compare the ratio of uncorrectable amplitude error to phase error for rotation error.
It is evident that even for modest values of $J=5/2, 7/2, \text{ and } 9/2$, the bit-flip error rate for the logical qubit is significantly suppressed compared to phase-flip errors.

The proposed encoding can be considered a generalized version of the Shor code,
\begin{equation}
\begin{aligned}
\ket{0}&=\frac{1}{\sqrt{8}}\left(\ket{\uparrow}^{\otimes 2J+1 }+\ket{\downarrow}^{\otimes 2J+1 }\right)^{\otimes 3}\\
\ket{1}&=\frac{1}{\sqrt{8}}\left(\ket{\uparrow}^{\otimes 2J+1 }-\ket{\downarrow}^{\otimes 2J+1 }\right)^{\otimes 3}\\
\end{aligned}
\label{eq:extended_shor_code }
\end{equation} 
For the Shor code \cite{shorcode}, the inner encoding protects against bit-flip errors and the outer encoding protects against phase-flip errors. 
In our case, the inner layer protection originates from the encoding of the qubit in the spin-$J$ qudit, $\ket{\uparrow}^{\otimes 2J+1 } = \ket{J,J}$,  $\ket{\downarrow}^{\otimes 2J+1 } = \ket{J,-J}$.

\vspace{0.7cm}
\tocless{\section{Universal gate set and Rank-Preserving CNOT gate }}
{\label{sec:universal_gate_set}}
In this section, we establish a set of universal fault-tolerant operations for spin-cat qubits.  As discussed above, similar to~\cite{Aliferis2008fault,puri2020bias}, our strategy is to first correct for the dominant errors by encoding the biased qubit in a repetition code $\mathcal{C}_1$.
 After performing error correction corresponding to code $\mathcal{C}_1$, we obtain a logical qubit with reduced (but less biased) effective errors. We can then achieve FTQC by employing another level of concatenation using a generic  CSS code $\mathcal{C}_2$, as long as the effective noise strength is below the threshold of the code $\mathcal{C}_2$.

 To construct the universal gate sets, we target the following  physical level gates,
\begin{equation}
    \{\mathcal{P}_{\ket{0}},\mathcal{P}_{\ket{+}},\mathcal{M}_{X},\mathcal{M}_{Z},\mathrm{CNOT},ZZ(\theta),X,Y,Z\}.
    \label{eq:physical_level_gates}
\end{equation}
{Here $\mathcal{P}$ denotes state preparation, and $\mathcal{M}$ represents the measurement operators.}
{We require these spin-cat qubit operations to be ``rank-preserving" so that they do not convert correctable errors into uncorrectable ones. Using this gate set,} one can construct the following {logical} universal gate set {for} $\mathcal{C}_1$, 
\begin{equation}
     \{\mathcal{P}_{\ket{0}_{\mathrm{L}}},\mathcal{P}_{\ket{+}_{\mathrm{L}}},\mathcal{M}_{X_{\mathrm{L}}},\mathcal{M}_{Z_{\mathrm{L}}},\mathrm{CNOT}_{\mathrm{L}}\}\cup \{\mathcal{P}_{\ket{i}_{\mathrm{L}}}, \mathcal{P}_{\ket{T}_{\mathrm{L}}}\}.
      \label{eq:logical_level_gates}
\end{equation}
  To prepare the magic states $\mathcal{P}_{\ket{i}_{\mathrm{L}}}, \mathcal{P}_{\ket{T}_{\mathrm{L}}}$, we can utilize rank-preserving $ZZ(\theta)$ at the physical level, similar to the bias-preserving case of qubits~\cite{zzPoulin} and cat codes~\cite{puri2020bias}.
  { The gate set given in  \cref{eq:logical_level_gates}} {has been previously studied} {\color{red}}as a possible gate set when there is a significant bias between different noise channels. 
 {For example, when we have a significantly large probability of dephasing noise compared to the bit flip noise as studied in \cite{Aliferis2008fault}.
 The studies in \cite{Aliferis2008fault} show that for a biased noise, this gate set gives a better threshold and overhead compared to the other gate set.
The threshold is improved by a factor of $5$ for the gate set in \cref{eq:logical_level_gates} for a biased noise compared to the unbiased noise \cite{Aliferis2008fault}. 
Also, the studies in \cite{Aliferis2008fault,puri2020bias}, showed that there is a significant reduction in the overhead for the gate set for a biased noise.}
\vspace{0.7cm}
\tocless{\subsection{Single qubit gates}}  

To ensure fault tolerance, a gate $U$ must not turn correctable errors into uncorrectable errors in a specific level of encoding, i.e., we require that
\begin{equation}
U \mathcal{E}_K U^{\dagger} \in \mathcal{E}_K,
\label{eq:fault_tolerant_condition}
\end{equation}
where $\mathcal{E}_K$ represents the set of correctable errors for the spin-cat encoding as defined in \cref{eq:error_channel_basis}.
Further, to prevent the propagation of correctable errors into uncorrectable ones during subsequent computations, the gates $U$ should act on states for which an error has occurred in the same manner as they act on states within the logical subspace.  
Specifically, these gates must exhibit identical behavior whether the states are in the cat subspace or the kitten subspace with $m>1$, the subspace corresponding to amplitude damping errors. 

 By building the gates $U$ in the universal gate set using operations solely from the spin-$J$ representations of $\mathrm{SU}(2)$, we can guarantee the condition in \cref{eq:fault_tolerant_condition}. To see this, recall the definition of spherical tensor operators:
\begin{equation}
    U T_{q}^{(k)} U^{\dagger}= \sum_{-k\leq q'\leq k} D_{q,q'}T_{q'}^{(k)},
    \label{eq:spherical_tensor_operators}
\end{equation}
where $U=e^{-i \theta \hat{\mathbf{n}}.\mathbf{J}}$ is a spin-$J$ $\mathrm{SU}(2)$ rotation operator and 
\begin{equation}
    D_{q,q'}=\bra{k,M=q'} \exp(-i \theta \hat{n}.\mathbf{J}) \ket{k,M=q},
    \label{eq:Wigner_D_matrix}
\end{equation}
are the elements of Wigner $D$-matrices \cite{sakurai2014modern}.
As a result, $\mathrm{SU(2)}$ operators do not change the rank of spherical tensor operators.
Using the above relationships for the basis of errors introduced in \cref{eq:basis}, we get,
\begin{equation}
\begin{aligned}
    U S^{(k)}_q U^{\dagger}&= \sum_{q'}\left( g_{q,q'} S^{(k)}_{q'}+\widetilde{g}_{q,q'}A^{(k)}_{q'}\right)\\
     U A^{(k)}_{q} U^{\dagger}&=\sum_{q'} \left( h_{q,q'} S^{(k)}_{q'}+\widetilde{h}_{q,q'}A^{(k)}_{q'} \right)
\end{aligned}
\end{equation}
where the coefficients $\{g_{q,q'},\widetilde{g}_{q,q'},h_{q,q'},\widetilde{h}_{q,q'}\}$ are given in \cref{sec:coefficients_rotation}.
Therefore, the SU(2) rotations do not change the rank of the error operators and obey the condition given in  \cref{eq:fault_tolerant_condition}.

The single-qubit Pauli gates for the qubit encoded in the spin-qudit  can be implemented using the following general $\mathrm{SU}(2)$ operations,
 \begin{equation}
 \label{eq:logical_paulis}
 \begin{aligned}
 X&=\exp(-i\pi J_x),\\
 Y&=\exp(-i\pi J_y),\\
 Z&=\exp(-i\pi J_z).
 \end{aligned}
 \end{equation} 
 These are easily implemented by Larmor precession of the spin.
 
In contrast, and critically, the Hadamard gate $H$ for the spin-cat encoding, defined by
\begin{equation}
    \begin{aligned}
    H\ket{0}&=\ket{+}.\\
    H\ket{1}&=\ket{-},
    \label{eq:Hadamard_gate}
    \end{aligned}
\end{equation}
{\em cannot} be  achieved by $\mathrm{SU}(2)$ operations alone. To see this, note that an SU(2) rotation preserves the projection of the spin onto a rotated axis.  As $\ket{0}$ and $\ket{1}$ are spin coherent states (so-called ``stretched states"), an SU(2) rotation cannot be used to prepare a cat state, which is a superposition of spin coherent states. 
{(The overlap of states generated by SU(2) operators acting on an eigenstate along $z$ and the cat state cannot exceed  $0.5$.)  The action of an SU(2) operator takes an eigenstate along $z$ to an eigenstate along a rotated axis.  The cat state is not an eigenstate of projection of angular momentum along any axis, and thus it cannot be mapped from $\ket{0}=\ket{J,J}$ by an SU(2) operation.
}
Therefore, 
\begin{equation}
    H \mathcal{E}_K H^{\dagger}\not \in\mathcal{E}_K.
\end{equation}
The essential feature of our protocol is to circumvent this restriction by using ancilla qubits and rank-preserving CNOT gates to effectively apply a Hadamard gate that preserves the set of correctable errors  {which is described in detail in \cref{sec:Hadamard_gate}}.\\

\tocless{\subsection{Rank-preserving CNOT gate}}
{\label{sec:CNOT_gate}}

In this section, we develop a rank-preserving CNOT gate, the key ingredient to realize the universal gate set, using only $\mathrm{SU}(2)$ operations. For concreteness, we provide a detailed protocol {based on the platform of neutral-atom quantum computing~\cite{Brennen1999, Deutsch2000, Jaksch_gate_2000, Saffman_review_2016_Rydberg,Browaeys_review_2020_Quantum}, which has shown increasing promise for scalable FTQC~\cite{Lukin_Nature_2022,Cong_Lukin_QEC_Rydberg_PRX_2022,Singh_Bernien_2022_dual_array,Saffman_Nature_2022,Ebadi_Lukin_2022_MIS_optimization}. 
In particular, we consider $^{87}$Sr atoms, with a spin-qudit encoded in the nuclear spin  $I=9/2$, providing a qudit with $d=10$ levels~{\cite{omanakuttan2021quantum,omanakuttan2022qudit}}.  When in the ground electronic state,} the weak coupling to the environment and resilience to other background noise makes the nuclear spin an ideal candidate for quantum information processing~\cite{barnes2021assembly,PhysRevLett.101.170504,daley2011quantum,PhysRevLett.98.070501}.  

Note, when considering the physical spins of atoms, in standard notation $I$ is the nuclear spin, $J$ is the total angular momentum of the electrons, and $F$ is the total electronic angular momentum plus nuclear spin. Our qudit is encoded in spin $I$ in the electronic ground state with $J=0$ for $^{87}$Sr, so that $F=I=9/2$.  
In this section, the spin angular momentum in which we encode the qudit is $\mathbf{F}$.  In the other sections of this article, we use $\mathbf{J}$ to denote a generic spin, without reference to its physical encoding.

We target a CNOT gate for the spin-cat encoding that operates the same for all kitten states. 
As discussed above (see \cref{eq:projectors}), we divide the qudit into ``left" and ``right" subspaces, with projectors onto them $\Pi_{\overline{0}}$ and $\Pi_{\overline{1}}$ respectively.  
The gate is formally given as,
\begin{equation} \label{eq:cnot_def}
    \mathrm{CNOT}= \Pi_{\overline{0}} \otimes \mathds{1}+\Pi_{\overline{1}} \otimes X,
\end{equation}
 where $X=\exp(-i\pi F_x)$.  
That is, we apply a $\pi$-rotation (NOT) to every kitten subspace of the target atom if the control atom is in the $\overline{1}$-subspace (the amplitude damped states of $\ket{1}$ we can correct), and the identity, if the control atom is in the $\overline{0}$-subspace (the amplitude damped states of $\ket{0}$ we can correct).  Clearly, if the amplitude damping takes an atom from the $\overline{0}$ to $\overline{1}$ space, or vice versa, the error cannot be corrected.

\begin{figure*}
   \subfloat[]{\includegraphics[width =1.1\columnwidth]{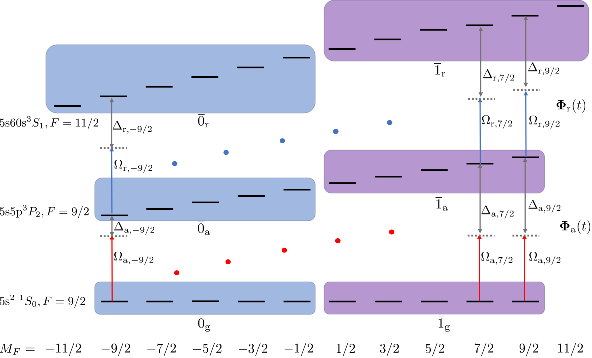} \label{fig:Fig_2_a}}
    \subfloat[]{\includegraphics[width =0.85\columnwidth]{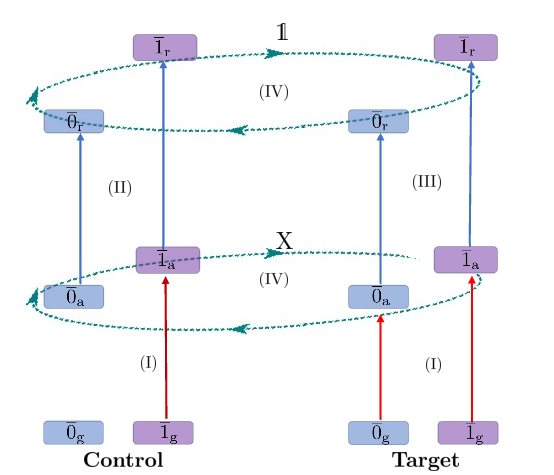}  \label{fig:Fig_2_b}}
\caption{ Protocol for implementing a rank-preserving CNOT-gate in neutral atomic $^{87}$Sr based of optimal control and the Rydberg blockade. The spin-cat qubit is encoded in the nuclear spin, ${I}=F=9/2$, in the electric ground state, $5\mathrm{s}^2 \; ^1{S}_0$.  (a) Detailed level diagram and protocol; (b) High-level schematic. When a gate is to be performed, the qudit is excited from the ground-state memory to the long-lived auxiliary metastable state,   $5\mathrm{s}5\mathrm{p}\;  ^3{P}_2$, ${F}=9/2$. Entangling interactions occur through excitation from the auxiliary state to the Rydberg state, $5\mathrm{s}60\mathrm{s} \; ^3S_1$, ${F}=11/2$. The error-correctable subspaces, $\overline{0}$ and $\overline{1}$, are represented by blue and purple colored boxes respectively, in the ground (g), auxiliary (a), and Rydberg (r) manifolds. The gate is performed in four steps.  Step I: Using quantum optimal control the population from the ground state is transferred to the auxiliary state while preserving coherence between magnetic sublevels. Each two-level resonance, $\ket{\mathrm{g},M_F}\rightarrow\ket{\mathrm{a},M_F}$, has a detuning $\Delta_{\mathrm{a},M_F}$ and Rabi frequency $\Omega_{\mathrm{a},M_F}$.  For the control atom, we only promote the population from the $\bar{1}$-subspace, whereas for the case of the target atom, we promote the population from both the $\overline{0}$ and $\overline{1}$ subspaces to the auxiliary state (see main text for details).  Step II: Using $\pi$-polarized light, local addressing, and quantum control, transfer the population from the auxiliary to Rydberg states only for the control atom. 
Step III: Apply the same pulse to the target atom.  Due to the Rydberg blockade, this population transfer only occurs when the control atom is in $\overline{0}$-subspace; for the  $\overline{1}$-subspace the population is otherwise blockaded.
Step IV: Using global rf-phase-modulated optimal control, we perform the SU(2) rotation $X=\exp(-i\pi F_x)$ in the auxiliary manifold and simultaneously the identity operator in the Rydberg manifold. The result is a CNOT gate -- if the control atom is in $\overline{1}$-subspace we apply an $X$ gate to the target atom if the control atom is in  $\overline{0}$-subspace we implement an identity operator $\mathds{1}$. Finally, we will transfer all the states back to the ground state by reversing steps III-I, thus implementing a rank-preserving CNOT gate for the spin-cat encoding.} 
\label{fig:fig_cnot}
\end{figure*}

The protocol for implementing this gate is presented in \cref{fig:Fig_2_b}. 
We note that this protocol requires individual addressing of the atoms.  In step I of the protocol, the population from the ground state memory is coherently transferred to an auxiliary state where it is more easily controlled.  In $^{87}$Sr, we utilize the auxiliary hyperfine state, $\ket{5\mathrm{s}5\mathrm{p};\; ^3{P}_2; \; F=9/2, M_F}$ with hyperfine quantum numbers $F=9/2, M_F$. This manifold possesses a large magnetic dipole moment and a long lifetime. For the control atom, only the population of $\overline{1}$-subspace is transferred to the auxiliary manifold, whereas for the target atom, the population from both $\overline{1}$-subspace and $\overline{0}$-subspace is transferred.
Both of these are facilitated by an effective $\pi$-pulse between the ground and the auxiliary states, which one can implement using quantum optimal control, as discussed below.

In step II, an effective $\pi$-pulse is applied on the control atom between the auxiliary and the Rydberg state.  In step III, we apply the same $\pi$-pulse on the target atom. Due to the Rydberg blockade, this population exchange only occurs when the control atom is in $\overline{0}$-subspace. If the state of the control atom is in $\overline{1}$-subspace, the population from the auxiliary state of the target atom is blockaded from transferring to the Rydberg state. 

Subsequently in step IV, using a global interaction and quantum optimal control, we {simultaneously} implement a $X=\exp(-i\pi F_x)$ rotation in the auxiliary manifold and an identity operator in the Rydberg manifold {of the target atom}.  The net effect is that if the control atom is in $\overline{1}$-subspace an $X$ gate has been applied to the target atom and if the control atom is in  $\overline{0}$-subspace the identity operator has been applied on the target. We transfer all the states back to the ground state by applying steps III-I in reverse order. The whole procedure implements the desired rank-preserving CNOT gate for the spin-cat encoding in \cref{eq:cnot_def}.

\begin{figure*}
    \subfloat[]{\includegraphics[width =0.66\columnwidth]{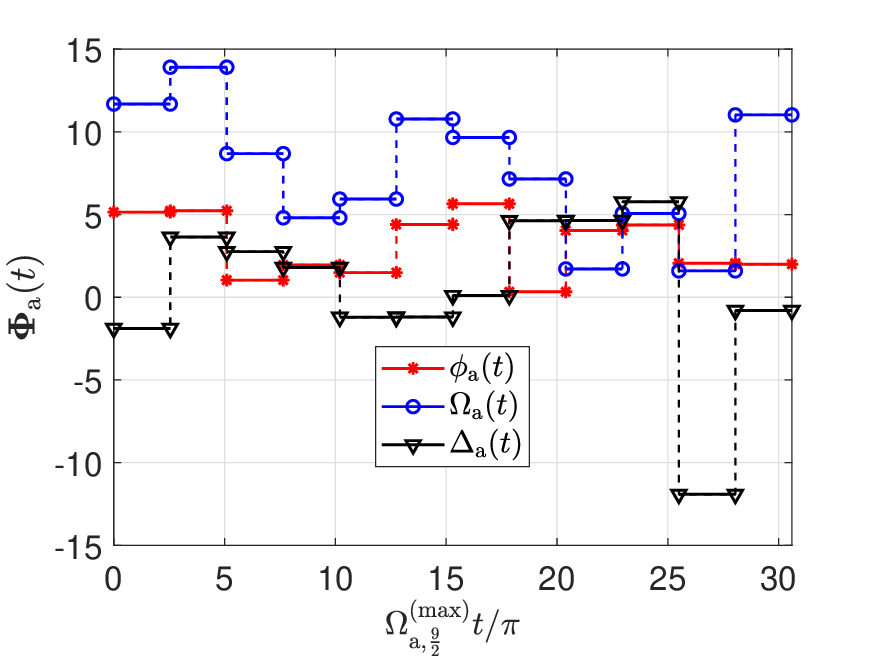} \label{fig:Fig_waveform_a}}
    \subfloat[]{\includegraphics[width=0.66\columnwidth]{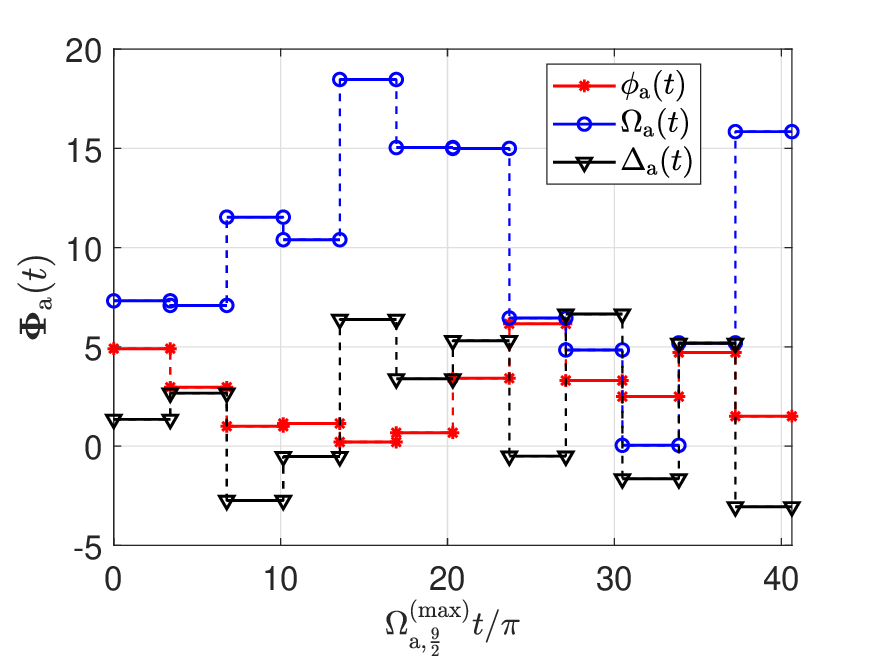}  \label{fig:Fig_waveform_b}}
        \subfloat[]{\includegraphics[width=0.66\columnwidth]{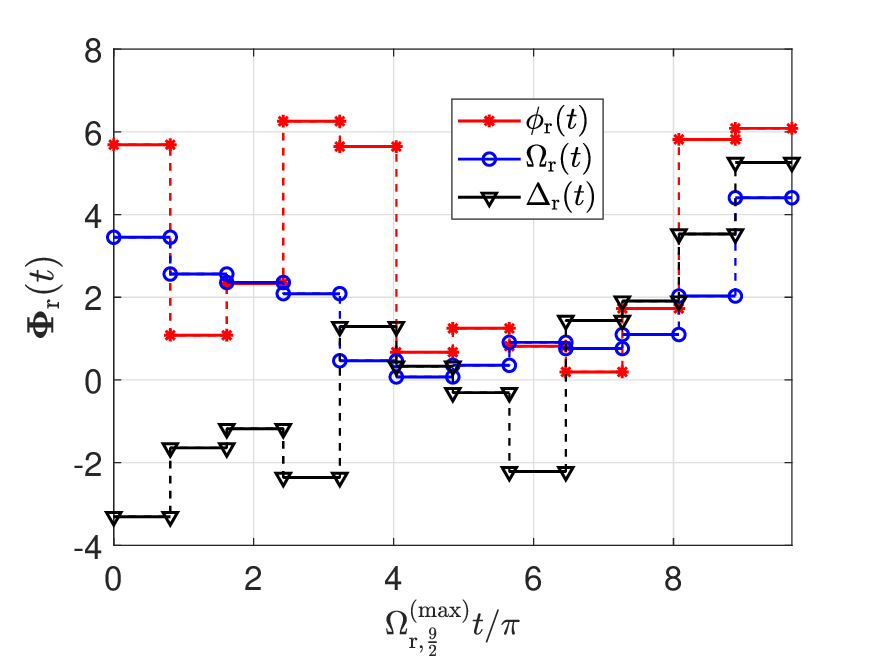}  \label{fig:Fig_waveform_c}}
    \caption{Examples of control waveforms that achieve the transfer of populations between spin manifolds while preserving the coherence between magnetic sublevels.  Based on Hamiltonian~\cref{eq:control_hamiltonian_cnot}, we modulate the lasers' amplitude, detuning, and phase, as piecewise constant functions of time.  Using the GRAPE optimal control we find the target isometries.    (a) The waveform that implements $V_{\mathrm{tar}}^{(\mathrm{C})}$, which transfer population from $\overline{1}_\mathrm{g}$-subspace to $\bar{1}_{\mathrm{a}}$-subspace while the population in the $\bar{0}_\mathrm{g}$-subspace is unchanged.
    (b) The waveform that implements $V_{\mathrm{tar}}^{(\mathrm{T})}$, which transfer population from $\overline{1}_\mathrm{g}$-subspace to $\bar{1}_{\mathrm{a}}$-subspace and $\overline{0}_g$-subspace to $\bar{0}_{\mathrm{a}}$-subspace .
    (c) The waveform that implements $V_{\mathrm{tar}}^{(\mathrm{Ryd})}$ that transfers the population from the auxiliary states to the Rydberg states.
  For all these three cases we divide the time into $12$ equal time steps.}
    \label{fig:transition_waveform}
\end{figure*}

In steps I and II of the rank-preserving CNOT gate, one needs to implement the transfer of population from the ground to the auxiliary manifold and from the auxiliary manifold to the Rydberg manifold, respectively.  This can be achieved by an effective $\pi$-pulse between these respective states and using quantum optimal control.
In both these cases we use the control Rabi Hamiltonian
\begin{equation}
    \begin{aligned}
        H_e(t)=& \sum_{M=-\frac{9}{2}}^{\frac{9}{2}} -\Delta_{e,M}(t)\ketbra{e,M}\\
     &+ \Omega_{e,M}(t)\left[e^{i\phi_{e}(t)}\sigma^+_{e,M}+\mathrm{h.c}\right].
     \label{eq:control_hamiltonian_cnot}
    \end{aligned}
\end{equation}
To simplify the notation we have denoted the two excited metastable manifolds by $e$, where $e=\mathrm{a}$ (auxiliary states) and $e=\mathrm{r}$ (Rydberg states).  Together with the ground-state manifold,
\begin{equation}
\begin{aligned}
    \ket{r,M}&\equiv\ket{5\text{s}60\text{s};\; ^3S_1;\hspace{0.1 cm} F=\frac{11}{2},M_F=M},\\
    \ket{a,M}&\equiv\ket{5\text{s}5\text{p};\; ^3P_2;\hspace{0.1 cm} F=\frac{9}{2},M_F=M},\\
    \ket{g,M}&\equiv\ket{5\text{s}^2;\; ^1S_0,;\hspace{0.1 cm} F=\frac{9}{2},M_F=M},
\end{aligned}    
\end{equation}
and
\begin{equation}
        \sigma^+_{e,M} \equiv \ketbra{e,M}{e',M}.
 \end{equation}
 where $e'=\mathrm{g}$ (for the interaction between the ground and auxiliary states) and $e'=\mathrm{a}$ (for the interaction between auxiliary and Rydberg states).
The control task is achieved by modulation of the amplitude, detuning, and phase of the exciting lasers. 
The time-dependent Rabi frequency and detuning are, 
 \begin{equation}
     \begin{aligned}
         \Omega_{e,M}(t)&=\mathcal{C}_{e,M}\Omega_{e}(t),\\
         \Delta_{e,M}(t)&=\Delta_{e}(t)+\delta_{e,M},
     \end{aligned}
 \end{equation}
 where $\mathcal{C}_{e,M}$ is the ratio of Clebsch-Gordan coefficients,
 \begin{equation}
    \mathcal{C}_{e,M}=\frac{\bra{F,M}\ket{1,0;F,M}}{\bra{F,\frac{9}{2}}\ket{1,0;F,\frac{9}{2}}}. 
 \end{equation}
$\Delta_{e}$ is the detuning, and $\delta_{e,M}$ is the additional detuning due to the relative Zeeman shift. 
 To implement the particular target unitary map interest ($U_{\mathrm{tar}}$) we consider modulation of the amplitude, detuning, and phase of the two lasers that drive the $\ket{g}\rightarrow\ket{a}$ transitions and the $\ket{a} \rightarrow \ket{r}$ transitions.  The GRAPE algorithm searches for the optimal control parameters $\Phi=\{\Omega_{e}(t), \Delta_{e}(t),\phi_{e}(t)\}$ that maximizes the fidelity with the target map $U_{\mathrm{tar}}$ 
 \begin{equation}
     \mathcal{F}[\Phi]=\frac{1}{{d}^2}\left| \trace\left\{ U_{\mathrm{tar}}^{\dagger}U[\Phi,T]\right\}\right|^2, 
 \end{equation}
where ${d}$ is the dimension of the qudit and $U[\Phi,T]=\mathcal{T}\left[\exp\left(-i\int_0^T H[\Phi(t)]dt\right)\right]$ is the solution to the time-dependent Schr\"{o}dinger equation. 

We consider partial isometries for our target maps. These have fewer constraints than unitary transformations and hence require fewer resources (time, bandwidth etc.).
For a system of dimension $d$, one can define a partial isometry as,
\begin{equation}
    V_{\mathrm{tar}}=\sum_{i=1}^{k}\ketbra{f_i}{e_i}
\end{equation}
where $1\leq k\leq d$ is the dimension of the partial isometry of interest and $\{\ket{e_i}\},\{\ket{f_i}\}$ are two orthonormal bases.
The unitary maps of interest then take the form, 
\begin{equation}
   U_{\mathrm{tar}}=V_{\mathrm{tar}}+V_{\perp}, 
\end{equation}
where $V_{\perp}$ acts on the orthogonal subspace, with dimension $d-k$.
To find the  control waveform that generates the partial isometry, one then optimizes the fidelity between the target isometry and the isometry generated using quantum control~\cite{pedersen2007fidelity}
\begin{equation}
\mathcal{F}_V[\Phi]=\left|\Tr\left(V^{\dagger}_{\text{tar}}V[\Phi,T]\right)\right|^2/k^2.
\label{eq:fidelity}
\end{equation}

For the case of the rank-preserving CNOT gate, one needs to implement three target isometries.
Firstly, on the control atom (C) we need to transfer the population from the $\overline{1}$-subspace of the ground manifold to that of the auxiliary manifold while keeping the population in the $\overline{0}$-subspace unchanged.  
The isometry we need to implement is,
\begin{equation}
    \begin{aligned}
        V_{\mathrm{tar}}^{(\mathrm{C})}&=\sum_{M=-\frac{9}{2}}^{-\frac{1}{2}}\ketbra{\mathrm{a},M}{\mathrm{a},M}+\sum_{M=\frac{1}{2}}^{\frac{9}{2}}\ketbra{\mathrm{a},M}{\mathrm{g},M}.
    \end{aligned}
\end{equation}
Secondly, we seek to transfer the entire population from the ground manifold to the auxiliary manifold on the target atom (T).  The isometry is
\begin{equation}
    \begin{aligned}
        V_{\mathrm{tar}}^{(\mathrm{T})}&=\sum_{M=-\frac{9}{2}}^{\frac{9}{2}}\ketbra{\mathrm{a},M}{\mathrm{g},M}.
    \end{aligned}
\end{equation}
Finally, we need to implement an isometry that transfers the population from the auxiliary manifold to the Rydberg manifold,
\begin{equation}
    \begin{aligned}
        V_{\mathrm{tar}}^{(\mathrm{Ryd})}&=\sum_{M=-\frac{9}{2}}^{\frac{9}{2}}\ketbra{\mathrm{r},M}{\mathrm{a},M}.
    \end{aligned}
\end{equation}
All three can be implemented using the Rabi Hamiltonian.

As a proof of principle, we numerically optimize a piece-wise constant waveform based on the well-known GRAPE algorithm for quantum optimal control~ \cite{Merkel2009,merkel2009quantum,jurdjevic1972control,goerz2015optimizing}.  
Example waveforms that implement the target isometries are given in \cref{fig:transition_waveform}. 
The total time required is $4\pi/\Omega_{\mathrm{rf}}$, where $\Omega_{\mathrm{rf}}$ is the rf-Larmor precession rate, chosen to be resonant with the Zeeman splitting in the auxiliary auxiliary manifold.  
To achieve high fidelity control, we have divided the time into $12$ equal time steps.  In practice, other parameterizations could be used to yield smoother waveforms if bandwidth is limited.
 
\begin{figure}[!htp]
    \centering
    \includegraphics[width =1\columnwidth]{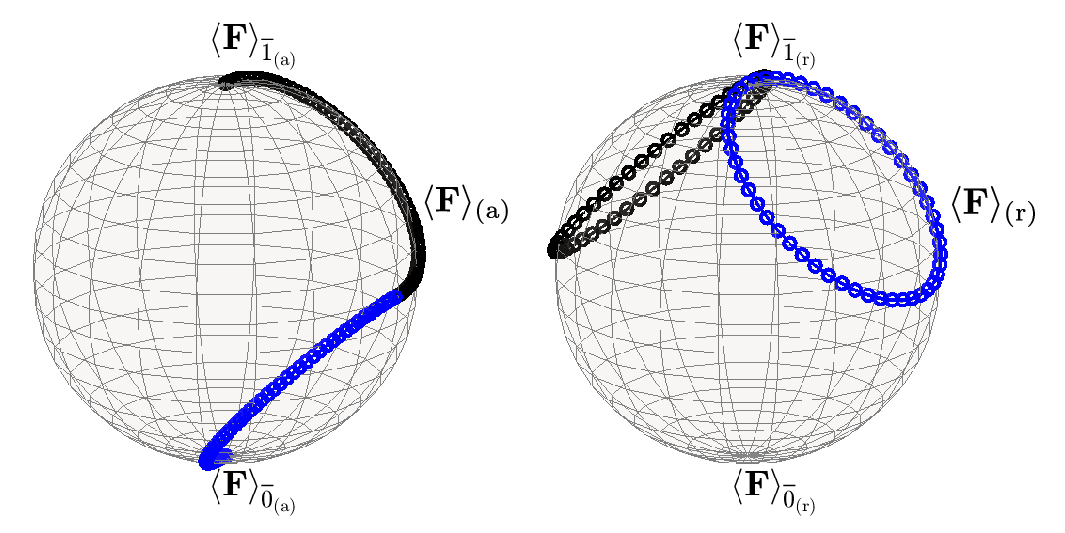} \label{fig:two_parameter_a}
    \caption{Evolution of the spin vector $\langle \mathbf{F} \rangle$ for the auxiliary (a) and Rydberg (r) manifolds resulting from rf-driven Larmor precession with time-varying phases. Optimal control is based on Hamiltonian~\cref{eq:Hamiltonians} for the piece-wise constant phases and total time $T_{\mathrm{tot}}=\sqrt{2}\pi/\Omega_{\mathrm{rf}}$. 
    The blue and black dots correspond to the first and second steps respectively (see text).
    An $X=\exp(-i\pi F_x)$ gate acts on the auxiliary manifold and transfers the population from $\bar{1}_{\mathrm{a}}$ to $\bar{0}_{\mathrm{a}}$ and vice-versa. However, for the   Rydberg manifold, the pulse sequence acts as an identity operator, and the population in the  $\bar{0}_{\mathrm{r}}$ and  $\bar{1}_{\mathrm{r}}$ subspaces remain unaffected.
    \label{fig:independent_evolution_of_excited_and_ground_state}}
\end{figure}

Another important ingredient for the rank-preserving CNOT gate in  \cref{fig:fig_cnot} is that we need to apply an rf-pulse that rotates the auxiliary $^3{P}_2$ state and the Rydberg $^3{S}_1$ state differently. For the case of the rank-preserving CNOT gate, one needs to implement an $X$ gate in the auxiliary manifold and identity in the Rydberg manifold.
This can be achieved because of the different magnetic $g$-factors of the two spin manifolds. 
For our specific choice of Rydberg manifold and auxiliary manifold $g_{\mathrm{r}}/g_{\mathrm{a}} \approx 2$~\cite{urech2023single}. The Hamiltonian describing Larmor precession in each of the excited manifolds, driven by an rf-magnetic field oscillating at frequency $\omega$, in the presence of a basis magnetic field is then
\begin{equation}
\begin{aligned}
H_a&=\Omega_{\mathrm{rf}} \left[\cos(\omega t+\phi) F_x +\sin (\omega t+\phi) F_y\right] +\omega_0 F_z,\\
H_r&=2\Omega_{\mathrm{rf}} \left[\cos(\omega t+\phi) F_x +\sin (\omega t+\phi) F_y\right] +2\omega_0 F_z.
\end{aligned}
\end{equation}
Here $\Omega_{\mathrm{rf}}$ is the Larmor precession frequency and $\omega_0$ is the Zeeman shift induced by the bias B-field in the $^3P_2$ auxiliary manifold.  The spin angular moment operators act in the respective manifolds.  Going to the rotation frame of the rf-oscillation, using the unitary operator $U=\exp(-i\omega t F_z)$, and choosing the rf-frequency to be off-resonant with $\omega=4/3\omega_0$, gives
\begin{equation}
\label{eq:Hamiltonians}
\begin{aligned}
H_a^{\mathrm{rot}}&=\Omega_{\mathrm{rf}}\left[\cos (\phi)F_x +\sin (\phi )F_y\right]-\frac{1}{3}\omega_0 F_z.\\
H_r^{\mathrm{rot}}&=2\Omega_{\mathrm{rf}} \left[\cos (\phi)F_x +\sin (\phi)F_y\right]+\frac{2}{3}\omega_0 F_z.
\end{aligned}
\end{equation}
Because of the finite detuning, the total Larmor precession frequency in the auxiliary and Rydberg manifold is then
\begin{equation} 
\begin{aligned}
   \Omega_{\mathrm{a}}&=\sqrt{\Omega_{\mathrm{rf}}^2+\frac{\omega_0^2}{9}}, \\
    \Omega_{\mathrm{r}}&=\sqrt{4\Omega_{\mathrm{rf}}^2+\frac{4\omega_0^2}{9}}= 2\Omega_{\mathrm{a}}.
\end{aligned}
    \label{eq:effective_Rabi_frequency}
\end{equation}
Since the total Larmor frequency of the auxiliary auxiliary and Rydberg manifolds are different, one can use optimization techniques such composite pulses \cite{levitt1986composite} or quantum optimal control \cite{Merkel2009,goerz2015optimizing} to implement separate unitaries in the auxiliary and Rydberg manifold. 

For example, when $\Omega_{\mathrm{rf}}=\omega_0/3$ using optimal control one achieves an $X$ gate in the auxiliary manifold and the identity in the Rydberg manifold by taking the phase to be a piece-wise constant function time, corresponding to a series of rf-pulses, and a total time, $T_{\mathrm{tot}}=\frac{\sqrt{2}\pi}{\Omega_{\mathrm{rf}}}$.
The resultant dynamics for the auxiliary and Rydberg manifold are given in \cref{fig:independent_evolution_of_excited_and_ground_state}. Since the optimization is purely geometric in nature the same pulse schemes work for any value of the spin as long as the $g$-factors have this ratio.
For further details on the optimization see \cref{sec:Rotating_the_ground_and_excited_manifold}.

The protocol described above can be generalized for other entangling gates.
One can optimize rf-phases in \cref{eq:Hamiltonians} to implement the identity operator in the Rydberg manifold and $R(\theta)=\exp(-i\theta \hat{\bm{n}}.\mathbf{F})$, an SU(2) operator, in the auxiliary manifold. 
Thus one can implement the gate $ZZ(\theta)=\exp{-i\theta Z\otimes Z}$ with any angle $\theta$, up to local $Z$ rotations, for the spin-cat qubits.

 \vspace{0.7cm}
\tocless{\subsection{State preparation and Measurement}}{
\label{sec:state_preperation_and_measurement}}
To complete the universal gate set, one needs to implement the state preparation and measurement at the physical level given in \cref{eq:physical_level_gates}. 
$\mathcal{P}_{\ket{0}}$, which is the preparation of the spin coherent state can be achieved with high fidelity using optical pumping  \cite{chow2022high}.
Also, $\mathcal{M}_Z$, which is the measurement in the $\ket{F, M_F}$ basis can be achieved with high fidelity in principle~\cite{barnes2022assembly,ringbauer2021universal}.
However, $\mathcal{P}_{\ket{+}}$ and  $\mathcal{M}_X$ are not straightforward to implement without an SU(2) Hadamard gate.  We describe here new approaches unique to spin-cat encoding and the rank-preserving CNOT gate.
\vspace{0.7cm}
\tocless{\subsubsection{Preparation of the spin-cat state}
\label{subsubsec:Preperation_of_the_cat_state}}
We can generate the spin-cat state $\ket{+}$ using multiple approaches. For example, one can use quantum optimal control by considering the controllable Hamiltonian
\begin{equation}
H(t)=\Omega_{\mathrm{rf}} \left( \cos \phi(t) F_x+\sin \phi(t)  F_y\right)+\beta F_z^2.
\label{eq:Control_Hamiltonian}
\end{equation}
This can be implemented in atomic systems using a combination of tensor light shifts and rf rotations~\cite{Paul_experiment_Cs_2007}.  
For the specific case of $^{87}$Sr, we have previously studied how this can be implemented with high fidelity through the tensor light shift imparted on the ground-electronic state nuclear spin~\cite{omanakuttan2021quantum}. Using quantum optimal control protocols one can generate the state $\ket{+}$ from an initial state $\ket{F, M_F=F}$.

The light-shift will also be accompanied by decoherence to photon scattering and optical pumping.
We study this in \cref{sec:Ratio_optical_pumping_errors} to calculate  the fidelity for the state preparation,
\begin{equation}
    \mathcal{F}_{\mathrm{state}}=\bra{+} \rho\ket{+}.
\end{equation}
For the particular choice of $^{87}$Sr, we find the fidelity for quantum optimal control is $\mathcal{F}_{\mathrm{state}}=0.9998$.

\tocless{\subsubsection{Measurement of $X$}}

\begin{figure}
\centering
    \includegraphics[width=\columnwidth]{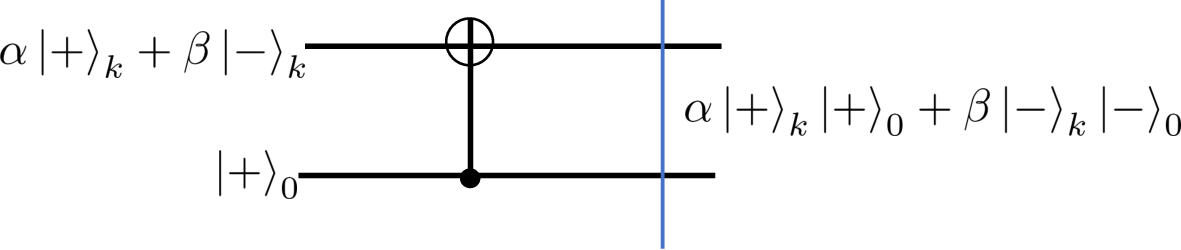}
\caption{Circuit diagram implementing $\mathcal{M}_X$. 
Consider an initial state $\alpha\ket{+}_k+\beta\ket{-}_k$, where $0\leq k \leq \lfloor \frac{2J-1}{2}\rfloor$, 
The action of the CNOT gate  for an ancilla state $\ket{+}_0\equiv\ket{+}$ gives us the state, $\alpha \ket{+}_k \ket{+}+\beta \ket{-}_k\ket{-}$, thus to identify whether the state is in $\ket{+}_k$ or $\ket{-}_k$, we need to measure whether the ancilla is in $\ket{+}_0$ or $\ket{-}_0$.
One can achieve this using a destructive measurement, (for more details see \cref{eq:X_measurement}). }
\label{fig:X_measurement}
\end{figure}
To measure the $X$ operator ($\mathcal{M}_X$), we need to identify whether the state is in $\ket{+}_k$ or $\ket{-}_k$ for $0\leq k \leq \lfloor \frac{2J-1}{2}\rfloor$. 
We cannot implement the $X$ measurement fault-tolerantly by applying a  Hadamard followed by measuring in the computational basis since Hadamard is not an SU(2) rank-preserving gate. 
{To surmount these challenges, similar to~\cite{puri2019stabilized}, we use an ancilla-assisted measurement protocol, where measurement errors will lead to syndrome errors without disturbing the encoded data.}  Hence, we implement the $X$-measurement by adding an ancilla qubit in the spin-cat state $\ket{+}_0$, applying a CNOT gate, and then destructively measuring the ancilla.
Since the ancilla is measured destructively and discarded, we do not need to implement the $X$-measurement using rank-preserving operators.

The circuit diagram which implements the measurement is shown in \cref{fig:X_measurement}. After the application of the CNOT gate, the joint state of the system is $\alpha \ket{+}_k \ket{+}_0+\beta \ket{-}_k\ket{-}_0$. 
Measuring whether the ancilla is in $\ket{+}_0$ or $\ket{-}_0$ gives the value of $X$ on the data qubit. 
To {measure the ancilla in the} $\ket{\pm}_0$ basis, we use quantum optimal control techniques to implement the required transformation to the $\mathcal{M}_z$ basis using $\mathrm{SU}(d)$ optimal control.  We employ the control Hamiltonian in \cref{eq:Control_Hamiltonian} to implement  the isometry \cite{omanakuttan2022qudit},
\begin{equation}\label{eq:X_measurement}
V_{\mathrm{targ}}= \ket{F, M_F=F}\bra{+}+\ket{F, M_F=-F}\bra{-}.
\end{equation}
In practice, this operation will be accompanied by decoherence, and the actual map we implement may be written as
\begin{equation}
V= e^{-\int \mathcal{L}(t)dt} V(0),
\end{equation}
where
\begin{equation}
    V(0)=\ket{+}\bra{+}+\ket{-}\bra{-}.
\end{equation}
and $\mathcal{L}(t)$ is the Lindbladian. 
Thus the fidelity for the implementation of the isometry is defined as
\begin{equation}
    \mathcal{F}_{\mathrm{iso}}=\frac{1}{4}\lvert\Tr(V_{\mathrm{targ}}V^{\dagger})\rvert^2.
    \label{eq:accuray_equation}
\end{equation}
  As an example, we consider the effect of photon scattering and optical pumping that accompanies the tensor light shift.  In our simulation, we achieve fidelity of $\mathcal{F}_{\mathrm{iso}}=0.999$ for  $^{ 87}$Sr in the presence of optical pumping described above.

We have now constructed all the required operations at the level of the qubit encoded in the spin, as given in  \cref{eq:physical_level_gates}. We can use these operations {to implement a universal gate set on the spin-cat qubits and} to construct the error correction and logical operations of the $\mathcal{C}_1$  code \cite{Aliferis2008fault}. (See \cref{sec:implementing_the_logical_operator} for the implementation of logical operations in $\mathcal{C}_1$.)
{{In principle, one can implement all the gates in \cref{eq:physical_level_gates} with very high fidelity, however, in practice one needs to consider other experimental imperfections like the Doppler effect that could impact the fidelity of these gates. }}
 
Generalizations of rank-preserving gate sets at the physical level can reduce the circuit size for specific applications   \cite{guillaud2019repetition,PRXQuantum.4.030311}.
For example, we can easily generalize our construction of the CNOT gate in \cref{sec:CNOT_gate} to implement a Toffoli gate in spin systems as discussed in~\cref{sec:Toffoli_gate}. 
The scheme is similar to the $\mathrm{CCZ}$ gate implemented in \cite{Levine_Pichler_gate}. 
This utilizes the capability to move neutral atoms in tweezer arrays, arranging the nearest neighbors to interact via the Rydberg blockade, while leaving the next-to-nearest neighbors unaffected. With access to such a gate, similar to the recent development in the bosonic system \cite{guillaud2019repetition}, we can implement the following operations,
\begin{equation}
\{P_{\ket{\pm}},\mathcal{M}_{X},\mathrm{CNOT},\mathrm{Toffoli}\}.
    \label{eq:universal_gate_set_1}
\end{equation}
Such a gate set can be used to construct more efficient fault-tolerant logical-level operations.

\vspace{0.7cm}
\tocless{\section{Syndrome Measurement and error recovery}}{
\label{sec:syndrome_extraction}}

\begin{figure}
\centering
\includegraphics[width=0.8\columnwidth]{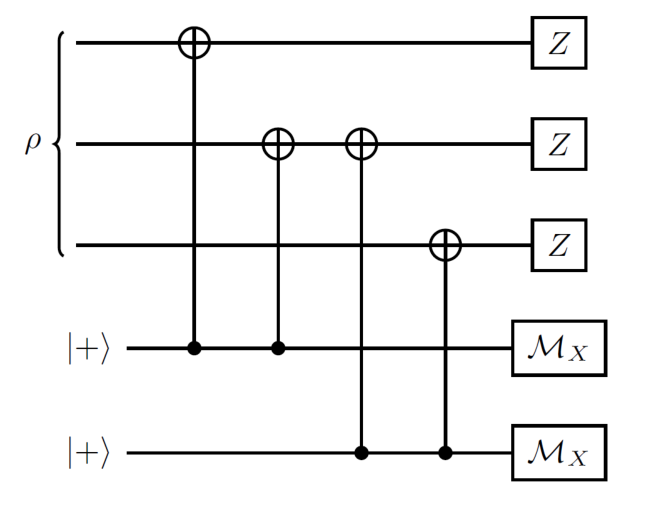}
\caption{Circuit for error correction of a phase error for a qubit encoded in $3$ spins. 
The error correction is achieved by measuring the syndromes $\{X_1X_2,X_2X_3\}$ followed by  $Z=\exp(-i\pi J_z)$ gate(s) according to the syndrome outcomes.
}
\label{fig:circuit_phase_error_correction}
\end{figure}

The design of error correction gadgets plays a major role in determining the threshold of tolerable noise and also the required overhead of fault-tolerant schemes mainly due to the fact that current fault-tolerant designs require many rounds of error correction to control the spread of errors. The standard method to perform an error recovery is to measure the syndromes to identify the errors and then correct the errors by applying an appropriate unitary operator. This is the approach we take to correct the phase errors. We use a repetition code of size $n$, capable of correcting up to $\lfloor (n-1)/2 \rfloor$ phase errors. In this case, the $(n-1)$ syndrome measurements for phase error correction are
\begin{equation}
\label{eq:syndrome_phase_detection}
\mathcal{S}_{\text{phase}}=\{X_1X_2,X_2X_3,\hdots, X_{n-1}X_n\}.
\end{equation}
These syndrome measurements can be implemented according to the standard circuits in \cref{fig:circuit_phase_error_correction} (for $n=3$)
using the universal operations described in \cref{sec:universal_gate_set}.

When the probability of phase errors is larger than amplitude errors in each spin, increasing the size of the repetition code $n$ can reduce the probability of logical phase errors. However, increasing $n$ will increase the probability of logical amplitude errors due to the increase in the number of the required CNOT gates for the syndrome circuits. Therefore we can choose the optimal $n$ that brings the two types of errors to the same level, determined by the noise threshold required by the outer CSS code $\mathcal{C}_2$.


For the case of amplitude damping, one approach to diagnose the syndrome is to perform nondestructive measurements to identify the amplitude errors, for example, by measuring $J_z^2$.  In practice this can be difficult to implement experimentally. (In this section and below we return to denote a generic spin $J$, without reference to a specific platform.) Instead, we take advantage of the cat encoding and the unique properties of our proposed CNOT gate to coherently apply the recovery map using fresh ancilla without performing any measurement. Our construction is a new example of measurement-free {quantum} error correction (MFQEC) \cite{PhysRevLett.117.130503,PhysRevA.97.012318,li2011recovery,premakumar2020measurement,PhysRevA.108.062426,,PRXQuantum.5.010333} motivated by the experimental constraints of spin systems.

\begin{figure}[!ht]
\subfloat[]{\includegraphics[width =0.45\columnwidth]{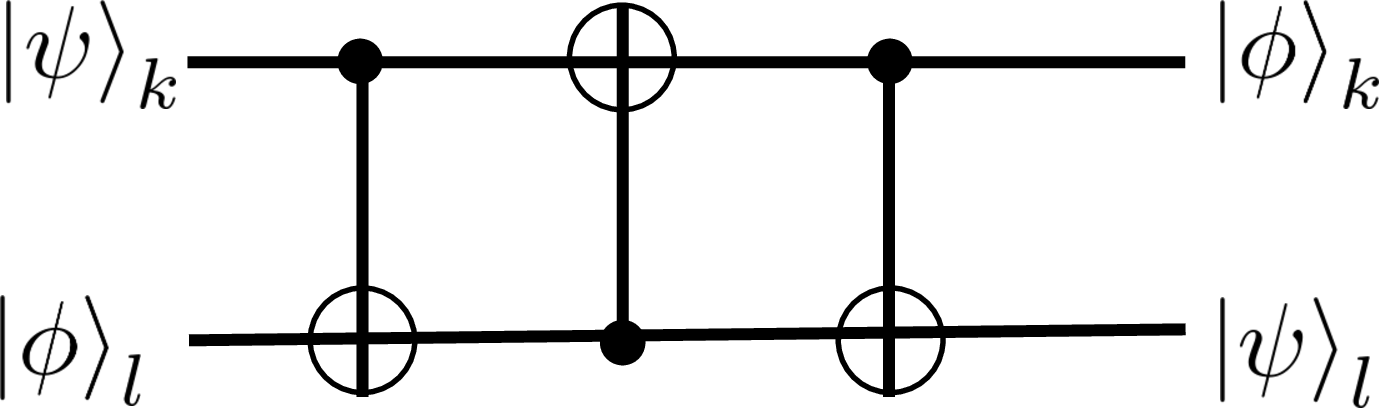} \label{fig:Fig_swap_a}} \hspace{0.5cm}
    \subfloat[]{\includegraphics[width=0.45\columnwidth]{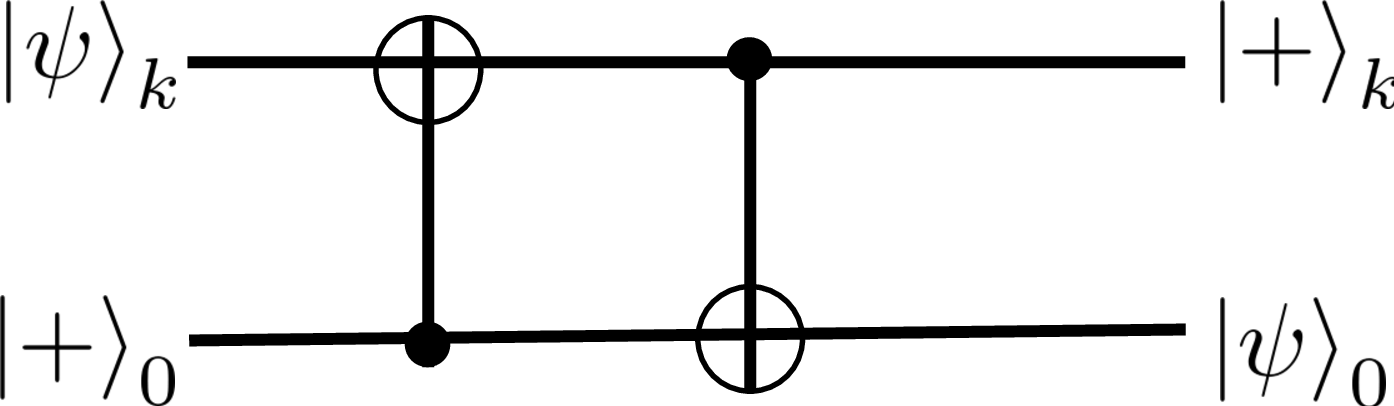}  \label{fig:Fig_swap_b}}
\caption{ (a) General circuit for swapping the state of the two qubits in two different kitten subspaces.
(b) The circuit that swaps the information between the data and ancilla, when the initial state of the ancilla state is $\ket{+}_0$.
}
\label{fig:SWAP_gate_qubit_1}
\end{figure}

To describe our proposed error recovery, we first observe that we can ``swap" the state of two qubits encoded in the kitten states. Let 
\begin{equation}
    \begin{aligned}
        \ket{\psi}_k&=\alpha \ket{+}_k+\beta\ket{-}_k,\\
        \ket{\phi}_l&=\gamma \ket{+}_l+\delta\ket{-}_l,\\
    \end{aligned}
\end{equation}
where $\alpha,\beta, \gamma, \text{ and } \delta$ are arbitrary complex amplitudes.
Three applications of our proposed CNOT gate, as shown in Fig.~\ref{fig:Fig_swap_a}, implement the following transformation (see \cref{sec:measurement_free_ec} for a proof):
\begin{equation}
    \begin{aligned}
        \ket{\psi}_k& \otimes \ket{\phi}_l \rightarrow \ket{\phi}_k \otimes \ket{\psi}_l 
    \end{aligned}
\end{equation}

We expect this construction which implements the swap of kitten states, to find applications beyond error correction, in particular in algorithmic subroutines native to qudits platforms, but in this work, we focus on its application in amplitude correction. We denote this gate by $V_s$.


If we replace one of the input states with a fixed cat state, $\ket{+}_0$, then the recovery circuit can be simplified to the circuit in Fig.~\ref{fig:Fig_swap_b}.  
Therefore amplitude errors can be corrected by consuming fresh ancilla qudits in the cat state, $\ket{+}_0$, and applying two CNOT gates. The operation coherently transfers the qubit that is damped at level $k$ back to level $0$, which is our encoded qubit.
In \cref{sec:measurement_free_ec}, we show that the action of this quantum channel, after tracing the extra subsystem, is exactly equivalent to a recovery channel implemented by measuring $J_z^2$ and then applying a unitary correction to transfer the state into the {$k=0$} subspace.

\vspace{0.7cm}
\tocless{\subsection{Error correction for Optical Pumping}}
To see how phase and amplitude error correction combines to correct any local errors, it is illuminating to describe the procedure for correcting a dominant noise channel in atomic systems, optical pumping. (The details of optical pumping are discussed in \cref{sec:Ratio_optical_pumping_errors}).
In particular, consider the example of absorption of a linear $\pi$-polarized laser photon, followed by the spontaneous emission of a circularly polarized $\sigma_{+}$ photon. 
This process results in mapping   $\ket{J,J}$ to $\ket{J,J-1}$ and also annihilating any amplitude in the state  $\ket{J,-J}$. On the cat states, this transformation can be re-written as,
\begin{equation}
    \begin{aligned}
        \ket{+} &\to \ket{J,J-1}= \frac{\ket{+}_1-\ket{-}_1}{\sqrt{2}},\\
          \ket{-} &\to -\ket{J,J-1}= -\frac{\ket{+}_1-\ket{-}_1}{\sqrt{2}}.
    \end{aligned}
\end{equation}
Consider an arbitrary logical state $\ket{\psi}=\alpha \ket{+}_\mathrm{L}+\beta \ket{-}_\mathrm{L}$. The action of the optical pumping on the first qudit gives
\begin{equation}
    \ket{\psi}\to \frac{\ket{+}_1-\ket{-}_1}{\sqrt{2}}\otimes \left(\alpha\ket{+}_0\ket{+}_0-\beta\ket{-}_0\ket{-}_0\right)\equiv\ket{\phi}.
\end{equation}
Now we can consider the states after the phase and amplitude error correction steps. (As these error correction steps commute with each other,  the order in which we perform  them is irrelevant.)
The phase error correction is specified by the syndromes $X_1X_2$ and $X_2X_3$. If we measure both the syndromes as $+1$, the state $\ket{\phi}$ collapses to,
\begin{equation}
    \ket{\phi} \to \alpha \ket{+}_1\ket{+}_0\ket{+}_0+\beta \ket{-}_1\ket{-}_0\ket{-}_0.
\end{equation}
If the syndrome measurement gives outcome $-1$ and $1$ for the syndrome $X_1X_2$ and $X_2X_3$ the state becomes,
\begin{equation}
    \ket{\phi} \to \alpha \ket{-}_1\ket{+}_0\ket{+}_0+\beta \ket{+}_1\ket{-}_0\ket{-}_0.
\end{equation}
Applying the correction unitary $Z_1$ corresponding to this syndrome yields
\begin{equation}
    \alpha \ket{+}_1\ket{+}_0\ket{+}_0+\beta \ket{-}_1\ket{-}_0\ket{-}_0\equiv\ket{\phi}_{\mathrm{ph}}.
    \label{eq:state_after_phase_correction}
\end{equation}
The same state is achieved after performing the correction for the other two possible syndromes. Thus the state after the phase error correction collapses to the state \cref{eq:state_after_phase_correction}. 

Next, we can apply measurement-free amplitude error correction by consuming three ancilla states $\ket{+}_0$, which gives,
\begin{equation}
\begin{aligned}
    V_s^{\otimes^ 3}\ket{\phi}_{\mathrm{ph}}\ket{+}_0\ket{+}_0\ket{+}_0
    &=\ket{+}_1\ket{+}_0\ket{+}_0 \otimes \ket{\psi}
\end{aligned}    
\end{equation}
Tracing out the first three subsystems yields the initial state $\ket{\psi}$ in the three ancilla subsystems. 
The error correction scheme developed here thus corrects the optical pumping errors.

This quantum error correction gadget is especially well suited to the neutral atom platform due to the ability to move atoms mid-circuit. Firstly, the swap gates are easy to implement as we can move individual ancillas and data atoms into a pairwise configuration to apply the CNOT gates parallelly. Secondly, at the end of the protocol, the ancilla atoms can be used as the new data atoms by simply moving them into the right positions.

\tocless{\section{logical CNOT gate and Fault-tolerant threshold}
\label{sec:Fault_tolerance_and_threshold}}

\begin{figure}
    \centering
    \includegraphics[width=\columnwidth]{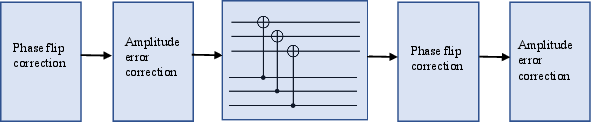}
    \caption{The error corrected {logical CNOT} gadget.
    The logical $\mathrm{CNOT}$ gate is implemented  by applying a physical CNOT gate between each qubit (encoded in the {spin}) of the control and target blocks transversely. 
    Error correction steps are performed before and after the logical CNOT. }
    \label{fig:fault_tolerant_cx_gadget}
\end{figure}

In this section, we provide lower bounds on the noise level that can be tolerated in our proposed spin-cat code, while still achieving fault-tolerant quantum computation. As discussed in \cref{sec:universal_gate_set}, to achieve fault-tolerance, we need to guarantee that the effective noise strength in our implementation of the logical gadgets of the inner code  $\mathcal{C}_1$, as specified by \cref{eq:logical_level_gates}, is below the noise threshold needed for the outer code $\mathcal{C}_2$ used in concatenation. 

In this concatenated scheme, the main source of error is the logical CNOT gate of $\mathcal{C}_1$, and hence, an upper bound on its failure probability will provide an estimate for the threshold of all $\mathcal{C}_2$ gadgets \cite{puri2020bias,Aliferis2008fault}. The logical CNOT gadget for the code $\mathcal{C}_1$ can be realized using transversal physical CNOT gates between two code blocks, accompanied by error correction procedures to correct phase and amplitude errors, which is illustrated in \cref{fig:fault_tolerant_cx_gadget}. For the sake of generality, we consider each logical CNOT gadget to consist of $r_1$ applications of phase error correction and $r_2$ applications of amplitude error correction. We define $r=r_1+r_2$ and denote the number of the data qudits in each code block by $n$. 
The recovery operation for phase error correction is determined by majority voting of the $r_1$ rounds of syndrome measurement.

We start by estimating the probability of dephasing errors. 
In this case, the analysis is similar to the analysis of biased cat qubits in bosonic systems \cite{puri2020bias}. 
Suppose each physical CNOT gate causes (independent) dephasing errors on the target and control qubits with probability $\epsilon$. 
During the application of {each} phase correction or amplitude correction procedures, every qudit is acted upon by at most two physical CNOTs.  
Hence, after $r_1$ repetition of phase corrections and $r_2$ repetition of amplitude corrections 
the probability of dephasing error on each qudit, in both the control and target block, will be at most $2r\epsilon$. 
After the implementation of error correction steps, the next step is to implement the transversal CNOTs between the control and target blocks of data qudits.
This operation can propagate phase errors from the target block to the control block. Therefore, after the action of the transversal CNOT gates,  the probability of dephasing error on each qubit of the target and control blocks is at most $2r\epsilon+\epsilon$ and $4r\epsilon+\epsilon$ respectively.

A logical error would occur if more than $(n+1)/2$ qubits are faulty in either the target or the control code blocks. 
Thus the upper bound on the logical phase error probability in the control and the target blocks can be  given as (keeping only the dominant term),
\begin{equation}
\begin{aligned}
\mathcal{\epsilon}_{\mathrm{target}}^{\mathrm{phase}}&\leq \binom{n}{\frac{n+1}{2}}(2r\epsilon+\epsilon)^{(n+1)/2},\\ \mathcal{\epsilon}_{\mathrm{control}}^{\mathrm{phase}}&\leq \binom{n}{\frac{n+1}{2}}(4r\epsilon+\epsilon)^{(n+1)/2}.
\end{aligned}
\end{equation}

To account for the possible errors in the syndrome measurements in the phase error correction step, we repeat measurements of $(n-1)$ syndromes in the control and the target blocks $r_1$  times and take the majority vote to apply error correction. 
A logical error happens if the syndrome is incorrect for at least $(r_1+1)/2$ rounds of this procedure. 
Each syndrome measurement requires two physical CNOT gates and we also need to account for state preparation and measurement errors used in each syndrome measurement, both of which can be performed with much higher accuracy compared to the rank-preserving CNOT gate. 
Also one needs to account for the dephasing error induced by the amplitude error correction following the phase error correction which has two physical CNOT gates.
Therefore the upper bound on the probability of a dephasing error in each syndrome bit is at most $6\epsilon$. 
As a result, the upper bound on the logical error for the syndrome measurement is given by (only keeping the dominant term):
\begin{equation}
    \mathcal{\epsilon}_{\mathrm{ec}}^{\mathrm{phase}}\leq 2(n-1)\binom{r_1}{\frac{r_1+1}{2}}(6\epsilon)^{\frac{r_1+1}{2}}.
    \label{eq:syndrome_error}
\end{equation}

Next, we establish an upper bound on the probability of logical errors resulting from amplitude errors on the control and target, just before the amplitude error correction step. An amplitude error on an individual qudit occurs when a minimum of $k_{\mathrm{max}} = (2J-1)/2$ jumps has taken place. 
This can be determined by summing the probabilities of $k_{\mathrm{max}}$ jumps, given a total of $s$ CNOT gates and is expressed as $q(s, k_{\mathrm{max}})$ as given in \cref{eq:amplitude_error_ideal} ($s = 2r$).
Following the error correction steps, the subsequent phase involves implementing transversal CNOT gates between the control and target blocks of data qudits.
This operation, however, has the potential to propagate amplitude errors from the control block to the target block. 
Consequently, after the application of transversal CNOT gates, the probability of amplitude errors on each qubit in the target and control blocks is bounded by
\begin{equation}
    \begin{aligned}
        \mathcal{\epsilon}_{\mathrm{target}}^{\mathrm{amp}}&\leq 2nq(s=2r,k_{\mathrm{max}})+nq(s=1,k_{\mathrm{max}}),\\ \mathcal{\epsilon}_{\mathrm{control}}^{\mathrm{amp}}&\leq nq(s=2r,k_{\mathrm{max}})+nq(s=1,k_{\mathrm{max}}).
    \end{aligned}
    \label{eq:blocks_amplitude_error}
\end{equation}

\begin{figure}
    \centering
    \includegraphics[width=0.8\columnwidth]{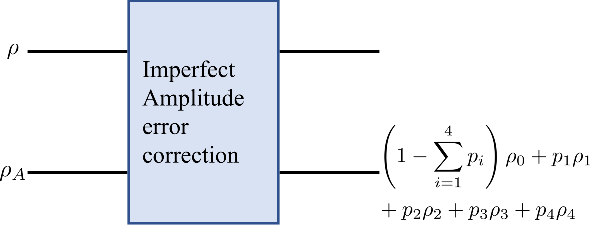}
    \caption{Imperfect amplitude error correction gadget. 
    There are two sources of imperfection one can associate with the amplitude error correction.
    The first one is a rotation error or optical pumping error occurring during the swapping approach to correct amplitude errors.
    The second one is due to imperfect preparation of the ancilla state, where ideally $\rho_A=\ket{+}_0$, however, in a non-ideal setting the ancilla can be in a mixture of $\ket{\pm}_i$ states where $i=\{0,1,2,3,4\}$, due to optical pumping or rotation error during the state preparation.
    For an ideal amplitude error correction, the final state lives in the $  \Pi_0=\ket{+}\bra{+}_0+\ket{-}_0\bra{-}_0$, whereas for a non-ideal setting, there is a small probability to be in other manifold $\Pi_l$. 
    The figure shows when the final state is in the $\Pi_i$ where $i=\{0,1,2,3,4\}$.}
    \label{fig:swapping_animation}
\end{figure}

Next, we provide upper bounds on the probability of logical error in the amplitude error correction procedure.
An ideal implementation of the swap protocol described in \cref{sec:syndrome_extraction}
would correct the amplitude errors by putting back the state into the cat manifold, defined as the support of the projector $\Pi_0$, where 
\begin{equation}
\Pi_l=\ket{+}_l\bra{+}_l+\ket{-}_l\bra{-}_l.
\end{equation}
Imperfect amplitude error correction may arise due to factors such as small random rotations during the swapping process intended for error correction, errors caused by optical pumping, or imperfections in ancilla preparation.
For the case of small random rotation errors and optical pumping, the error operators involve at most two amplitude jumps as discussed in \cref{subsec:Error_characterization}.
Similarly, as discussed in \cref{subsubsec:Preperation_of_the_cat_state}, 
optical pumping and random rotation errors can create at most two amplitude jumps during the preparation of the ancilla state. 
Thus the imperfect amplitude error correction can cause at most four amplitude jumps. This phenomenon is {conceptualized} in \cref{fig:swapping_animation}, where
the population in the cat manifold can leak to $\Pi_i$ for $i=\{1,2,3,4\}$ manifolds with probabilities $p_i$.

Errors in the preparation of the ancilla can in principle result in a  superposition of $\ket{+}_k$ states with $k\leq 4$ instead of $\ket{+}_0$. 
However, the amplitude error correction destroys any coherence between the cat and kitten subspaces, resulting in a mixed state in the cat manifold (see \cref{sec:measurement_free_ec}).
Hence, to find an upper bound on the success probability of amplitude correction, we only need to consider the probability of error in preparing  $\ket{+}_k$ states with $k\leq 4$, rather than an arbitrary state in that subspace.

We denote the failure probability of the amplitude error correction given that the ancilla states is in $\ket{+}_k$  by  $q(s,k_{\mathrm{max}}|k)$ where $s$ is the total number of CNOT gates before the application of error correction, and $k_{\mathrm{max}}$ is the minimum rank of the amplitude errors which create a logical error, i.e.  $k_{\mathrm{max}}=\lfloor (2J+1)/2\rfloor$ in our construction.
This probability can be calculated by adding the probabilities of cascades of single and two jumps that push the population from level $k$ to at least $k_{\mathrm{max}}$ level. 
Assuming the population only leaks to $\Pi_i$ for $i=\{1,2,3,4\}$ the logical error probability after $r_2$ rounds of amplitude error correction can be bounded by 
\begin{equation}
 \begin{aligned}
 \epsilon^{\mathrm{amp}} \leq r_2\left(\sum_{k=0}^4 q(s,k_{\mathrm{max}}|k)p_k\right),
    \end{aligned}
\end{equation}
where $p_0=1-\sum_{i=1}^4p_i$. 
(For a detailed calculation see \cref{sec:upper_bounds_amplitude}.)
As we have $2n$ total qudits, the logical error probability of the amplitude error correction 
blocks for the logical CNOT gate can be bounded by 
  \begin{equation}
 \begin{aligned} \mathcal{\epsilon}_{\mathrm{ec}}^{\mathrm{amp}}\leq 2n\epsilon^{\mathrm{amp}}.
    \end{aligned}
\end{equation}
Note that unlike phase error correction where the measurement is repeated $r_1$ many times and the correction is applied based on a majority vote of syndrome results, amplitude error correction does not involve direct measurement. Therefore repeated applications of amplitude error correction without a phase correction step in between do not provide extra error correction power.

Finally adding up all the probabilities of failures for the various components of the logical CNOT gate, yields an upper bound on its total logical error probability,
\begin{equation}
\begin{aligned}    \epsilon_{\mathrm{logical}}&\leq\mathcal{\epsilon}_{\mathrm{ec}}^{\mathrm{phase}}+\mathcal{\epsilon}_{\mathrm{control}}^{\mathrm{phase}}+\mathcal{\epsilon}_{\mathrm{target}}^{\mathrm{phase}}\\
&+\mathcal{\epsilon}^{\mathrm{amp}}_{\mathrm{ec}}+\mathcal{\epsilon}_{\mathrm{control}}^{\mathrm{amp}}+\mathcal{\epsilon}_{\mathrm{target}}^{\mathrm{amp}}.
\end{aligned}
    \label{eq:error_rate_total}
\end{equation}

To assess the improvement provided by our construction, we provide estimates of $\epsilon$ for various noise parameters that guarantee a logical error $ \epsilon_{\mathrm{logical}}$ below the threshold demanded by the CSS code   $\mathcal{C}_2$. For the CSS code $\mathcal{C}_2$ we use the fault-tolerant construction of \cite{Aliferis_css_code}, with a provable threshold of $\mathcal{\epsilon}_{\mathrm{CSS}}=0.67\times  10^{-3}$.

\begin{figure*}[!ht]
    \centering
     \subfloat[]{\includegraphics[width =1\columnwidth]{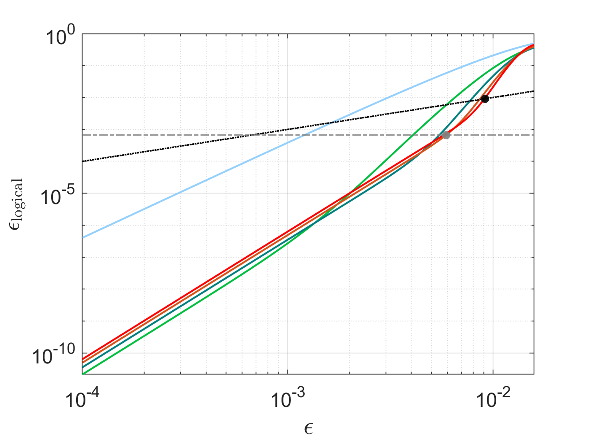} \label{fig:Fig_9_a}}
    \subfloat[]{\includegraphics[width =1\columnwidth]{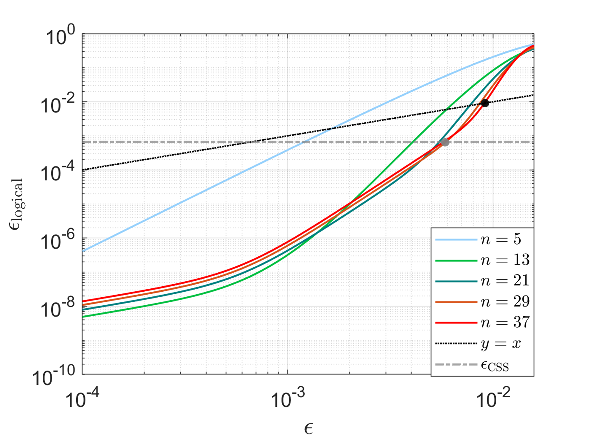}  \label{fig:Fig_9_b}}
    \caption{Logical error as a function of the physical level error (for details of the relation between phase error and amplitude error, see \cref{sec:Ratio_coherent_errors}) for the random rotation error for different value of $n$.
    Also, the threshold one needs to achieve CSS encoding in the second layer of concatenation is given for reference. 
   Figure (a) is for the case of $p_i=0$ for $i\neq 0$ and  figure (b) is for an imperfect ancilla state preparation with   $p_i=10^{-4}$ for $i\neq 0$. 
    We can see whether the swapping error ideal or non-ideal does not affect much except for very low noise and this in turn is because the contribution of the amplitude error is very low for the random rotation error.    
    The black circle shows the intersection of the logical error with $y=x$ line for the optimal case shown here and the gray circle shows the intersection of the $\mathcal{\epsilon}_{\mathrm{CSS}}$ with the logical error for the optimal case.
     The simulation is shown for $r_1=7$ and $r_2=1$.}
    \label{fig:coherent_error}
\end{figure*}
In \cref{fig:coherent_error} we present 
the case of the small rotation error for encoding a qubit in a qudit $J=9/2$ with $r_1=7$, $r_2=1$, and for different choices of $n$. 
The figure on the left assumes no leakage error in the ancilla state preparation, i.e.  $p_i=0$ for $i\neq 0$, and the figure on the right is for a leakage error of   $p_i=10^{-4}$ for $i\neq 0$.
As is evident in the figure, the logical error rates for scenarios with and without leakage error exhibit similar characteristics except for very low noise.
This is expected since for small rotation errors, the probability of amplitude error is exponentially suppressed as a function of $J$ compared to the phase errors, see \cref{fig:rotation_error_comparison} for more details. In particular, we find that for $n=21$, $r_1=7$, and $r_2=1$, the physical error $\mathcal{\epsilon}$ needed to achieve the targeted CSS threshold is less than $0.0054$. 

\begin{figure*}[!ht]
    \centering
    \subfloat[]{\includegraphics[width =1\columnwidth]{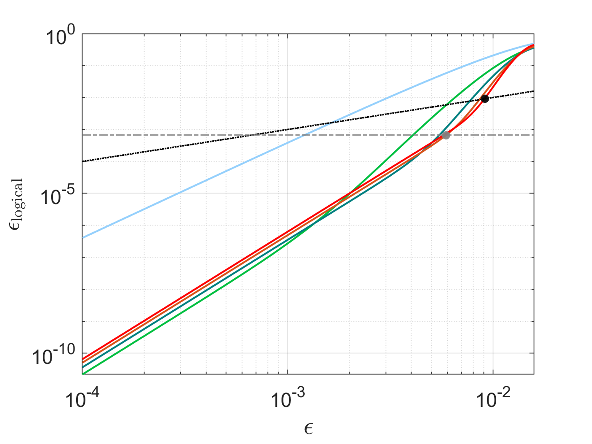} \label{fig:Fig_10_a}}
    \subfloat[]{\includegraphics[width =1\columnwidth]{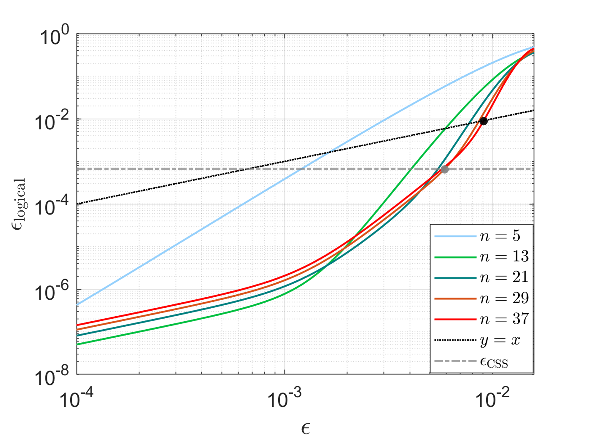}  \label{fig:Fig_10_b}}
    \caption{Logical error as a function of the physical level error (for details to the relation between phase error and amplitude error, see \cref{sec:Ratio_optical_pumping_errors}) for the optical pumping error for different value of $n$.
    The targeted threshold for the CSS code in the second layer of concatenation is given for reference. 
    Figure (a) is for the case of $p_i=0$ for $i\neq 0$ and  figure (b)  is for an imperfect ancilla state preparation with    $p_i=10^{-4}$ for $i\neq 0$. 
    We can see a significant change in the behavior depending on whether the amplitude error correction is ideal or not specifically in the low noise regime. 
    This in turn is due to the fact that for the case of optical pumping, as seen in \cref{sec:Ratio_optical_pumping_errors}, there is a significant contribution to the logical error from the amplitude errors.        
        The black circle shows the threshold value for the optimal value of $n$ and the gray circle shows the intersection of the $\mathcal{\epsilon}_{\mathrm{CSS}}$ with the logical error for the optimal value of $n$.
        The simulation are shown for $r_1=7$ and $r_2=1$.}
    \label{fig:optical_error}
\end{figure*}

Next, in \cref{fig:optical_error} we explore the impact of stronger photon scattering and optical pumping on the encoding of a qubit in a qudit with $J=9/2$ with $r_1=7$,  $r_2=1$, considering various choices of $n$. 
(For the case of $J=9/2$, we get $\alpha=0.0137$ and $\beta=0.2$ in \cref{eq:jump_opt_simplified} for stronger photon scattering and optical pumping.
Details of the noise model and parameters can be found in \cref{sec:Ratio_optical_pumping_errors}.) The left panel is the case with no leakage error $p_i=0$ for $i\neq 0$, while the right panel incorporates a leakage error with   $p_i=10^{-4}$ for $i\neq 0$.

As can be seen in the figure, for the ideal amplitude error correction the behavior of both the rotation error and case when photon scattering and optical pumping are stronger are very similar in nature.
However, when photon scattering and optical pumping are stronger, the imperfect ancilla preparation during amplitude error correction plays a more severe role in the overall logical error of the low noise regime. The competition between the error correction power of the gadget and the extra error due to the increased number of qudits needed to encode a logical qubit leads to identifying a ``sweet spot" that determines the optimal number of qudits needed to encode a logical qubit. In particular, we find that for $n=21$, $r_1=7$, and $r_2=1$, the physical error needed to achieve the targeted CSS threshold is $\mathcal{\epsilon} \leq  0.0053$.

As discussed in detail in \cref{sec:Ratio_coherent_errors}, the primary error source for the considered spin systems is the first-order angular momentum operators, stemming from potential unwanted magnetic fields.
Additionally, there are second-order terms in the angular momentum operators due to optical pumping \cite{deutsch2010quantum,omanakuttan2021quantum}. 
Despite this, the presence of extra levels in the qudit results in a logical error contribution from amplitude errors that is notably lower than that from phase errors. 
Thus the threshold behavior for both these error models only impacts the  low noise regimes.

\tocless{\section{Summary and Outlook} \label{sec:discussions_and_future_work}}

To achieve the full power of quantum computing, one needs to execute quantum algorithms on error-corrected logical qubits. However, meeting the demanding requirements for physical qubits and achieving low error rates, essential for error-corrected logical qubits, remains a significant challenge in current quantum implementations~\cite{knill2005quantum,raussendorf2007topological,svore2006noise,spedalieri2008latency}.
Recent advancements in noise-tailored error correction provide a promising avenue for achieving this by substantially alleviating the stringent demands of error-corrected logical qubits~\cite{Aliferis2008fault, zzPoulin, Grassl_erasure_1997_PRA, wu2022erasure, Sahay_Puri_biased_erasure_PRX_2023}.

In this article, we follow this direction and introduce a fault-tolerant quantum computation protocol by encoding a qubit into a spin system, with a spin larger than $J=1/2$.
The general scheme that we introduce in this work is applicable to a wide range of physical spins, including in semiconductors~\cite{Gross2021, gross2021hardware}, atomic ions~\cite{ringbauer2021universal, low2020practical}, neutral atoms~\cite{omanakuttan2021quantum, omanakuttan2022qudit, zache2023fermion}, molecules~\cite{castro2021optimal}, and superconducting systems~\cite{ozguler2022numerical, blok2021quantum}, where we have spin qudits that can be coherently controlled and entangled.

The specific encoding we consider in this article is the spin-cat encoding which draws inspiration from the cat-code encoding for continuous variable bosonic systems~\cite{puri2020bias,Aliferis2008fault}.  For this implementation we develop techniques to perform reliable computation in the presence of dominant noise in spin systems, taking advantage of natively available interactions. One key factor that distinguishes the spin-cat encoding from the other encodings of a qubit in a qudit is that the total Hilbert space of the spin-cat encoding decomposes into a direct sum of qubit subspace.
This induces the structure of a stabilizer code, a feature that plays a pivotal role in enabling fault-tolerant schemes for error correction.

Spherical SU(2) tensor operators provide a basis in which to characterize the error channels and identify the set of correctable errors. The dominant error sources for encoding a qubit in a spin are the rank-1 SU(2) rotations and the rank-2 tensors which can arise, e.g., from optical pumping between magnetic sublevels.  Our codes are constructed with these physical errors in mind. We use the concatenation scheme of~\cite{Aliferis2008fault} to perform fault-tolerant computation. 
In addition to using an inner repetition code that corrects phase errors, we correct for amplitude-damping errors by consuming fresh ancilla spins and performing measurement-free error correction natively for spin systems.

As a concrete application of our proposed scheme, we focus on the encoding of a qubit in the nuclear spin of $^{87}$Sr, characterized by a spin of $9/2$. In this scenario, we systematically build a universal gate set for fault-tolerant quantum computing, leveraging the available interaction mechanisms.
A pivotal element in the formulation of the physical-level gate is the rank-preserving CNOT gate. 
We elaborate on the implementation details of this gate, by taking advantage of the metastable states available in $^{87}$Sr and the well-known Rydberg blockade. In addition to the swap gadget that helps us correct amplitude errors, this CNOT gate is used in the construction of a universal gate set.

We also studied the threshold for fault-tolerant error correction and found that it is much higher than found in standard protocols of error correction with physical qubits, and it is similar to the threshold observed in bosonic cat-codes \cite{puri2020bias}.
As a result, our approach demonstrates a significant reduction in the required overhead and exhibits higher fault tolerance thresholds compared to conventional qubit-based techniques.

Our work represents another example of designing resource-efficient fault-tolerant schemes by taking advantage of the native noise characteristics of a given hardware. In contrast to the earliest work in quantum error correction where models were constructed for hypothetical qubits and generic noise models, efforts are being made to develop error correcting codes that are symbiotic with the control methods and noise structures of physical quantum systems \cite{Gross2021,omanakuttan2023multispin,puri2020bias,puri2019stabilized,Cong_Lukin_QEC_Rydberg_PRX_2022,gross2021hardware}. A related direction of research is to engineer qubit encodings with favorable noise properties~\cite{puri2017engineering,wu2022erasure}. This has been made possible because of the substantial experimental advances in quantum computing~\cite{acharya2022suppressing,ryan2022implementing,krinner2022realizing}.

In a similar vein, the structure of our protocol works well with spin systems and their control methods, regardless of the platform in which they are implemented. It is particularly well-suited for the neutral atom platform, where significant experimental advances have been achieved recently~\cite{Bluvstein_Lukin_2023_QEC_Logical,Lukin_Nature_2022,Saffman_Nature_2022}. We have previously explored the use of quantum optimal control of spin-$9/2$ nuclei in $^{87}$Sr atoms for arbitray single qudit gates~\cite{omanakuttan2021quantum} and two-qudit entangling gates~\cite{omanakuttan2022qudit}, where this protocol would be a natural fit. The unique capabilities of neutral atom platforms, such as reconfigurable connectivity and the ability to implement hundreds of parallel entangling gates~\cite{Bluvstein_Lukin_2023_QEC_Logical} would assist in the implementation of the fault-tolerant protocol we proposed here.

This work opens many directions for future research. One can extend the current protocol for the rank-preserving CNOT gate in neutral atoms to other, more experimental-friendly protocols.  Specifically, one can explore using the geometric phase approach \cite{Levine_Pichler_gate} or Rydberg dressing-based approaches \cite{mitra_martin_gate, schine2022long, martin2021molmer,mitra_martin_gate_two_photon}, typically used for entangling gates in qubits, to realize the rank-preserving CNOT gate.  
In addition, similar to continuous-variable cat encoding~\cite{puri2020bias}, the proposed gate set enables the use of other codes, including topological codes.
 Another direct extension is to develop gate sets to perform computation by encoding a qudit, rather than a qubit, into the large spin.

Lastly, while we focus on errors caused by random rotations and optical pumping in this paper. 
Another very important source of errors we didn't consider is leakage out of computational subspace, especially in the form of atom loss in neutral atom platforms. The conventional approach to circumvent these errors is to use leakage reduction units \cite{suchara2015leakage}. 
{In future work, we plan to address leakage errors using a Quantum non-demolition measurement to measure the presence of population in the computational subspace without destroying the coherence~\cite{Omanakuttan_future}. This measurement converts all leakage errors, including atom loss, into erasure errors which are easier to correct.}

\vspace{0.7cm}

\tocless{\section{Acknowledgments} \label{section:supplementary}}

This material is based upon work supported by NSF CAREER award No. CCF-2237356. Additional support by the Quantum Leap Challenge Institutes program (Grant No. 2016244) is acknowledged. The authors acknowledge fruitful discussions with Anupam Mitra, Tyler Thurtel, Austin Daniel, and Karthik Chinni. 
\clearpage

\onecolumngrid
\tableofcontents
\appendix

\twocolumngrid

\section{Small Rotation errors}
\label{sec:Ratio_coherent_errors}
\begin{figure*}[t]
    \centering
    \subfloat[]{\includegraphics[width =1\columnwidth]{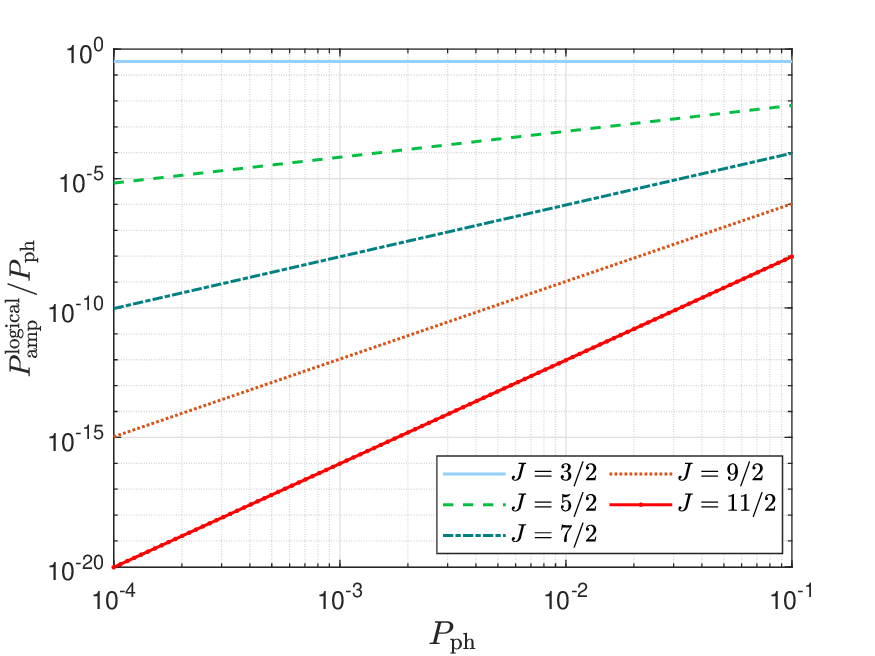} \label{fig:rot_error_a}}\hspace*{-1.9em}
    \subfloat[]{\includegraphics[width =1\columnwidth]{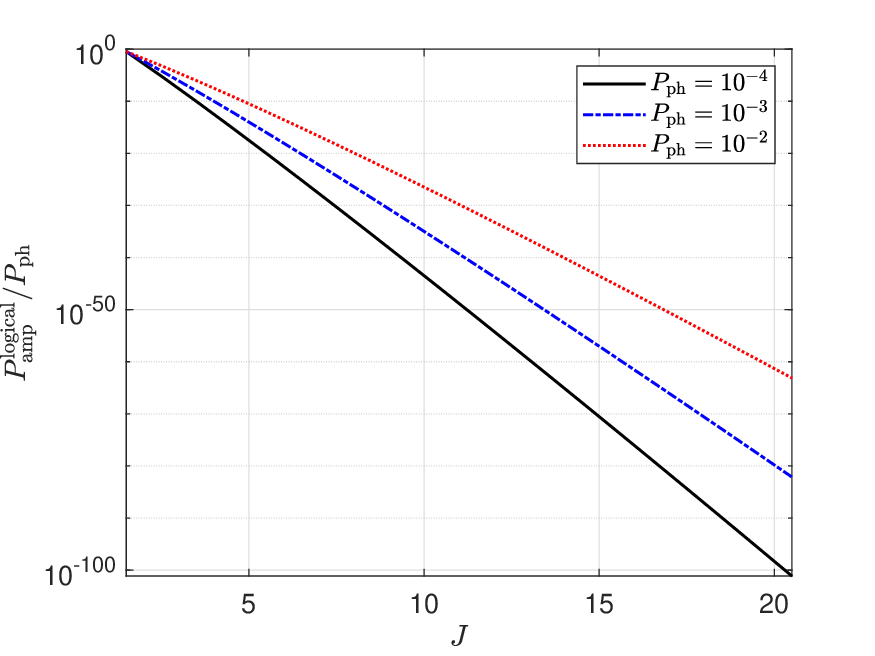}  \label{fig:rot_error_b}}
    \caption{Logical amplitude error probabilities due to rotation errors.
    (a) The ratio of logical amplitude error to phase error is given as a function of phase error.
    The probability of logical errors falls as the overall error rate decreases.
    A logical error occurs when we have $\lfloor (2J-1)/2\rfloor$ amplitude errors and thus as spin  $J$,  increases, the ratio decreases exponentially.
    However, for $J=3/2$, a single amplitude jump creates a logical error and thus the ratio of logical error to phase error is a constant equal to $1/2J$.
    (b)  The ratio of logical error probability due to amplitude errors to phase error for rotation error as a function of spin $J$. We can see that this ratio exhibits an exponential trend, and the logical error becomes negligible for sufficiently large values of $J$. 
  Consequently, fewer rounds of amplitude error correction will be needed as $J$ increases.}
    \label{fig:rotation_error_comparison}
\end{figure*}

A main source of decoherence for a qubit encoded in a spin is small random rotation errors \cite{gross2021hardware,Gross2021}.
As given in \cref{eq:error_prob_rotation}, for the spin-cat encoding the ratio of phase error to amplitude error decreases with spin $J$ as $1/J$.
However, for the spin-cat encoding, we need $\lfloor (2J-1)/2\rfloor$ amplitude errors/jumps for a logical error (logical amplitude error) to occur, such that these errors are not correctable by the amplitude error correction (a logical bit flip error for the encoding in \cref{eq:concat_spin_cat}). 
As such, we look at the probability of such logical amplitude errors in \cref{fig:rotation_error_comparison}.
In \cref{fig:rot_error_a} we show that for a spin $J$, the logical amplitude error decreases with phase error probability, and the decrease shows an exponential behavior with spin $J$.

To further illustrate the exponential suppression of the logical error arising from amplitude errors as a function of spin due to random rotation errors, in  \cref{fig:rot_error_b}, the ratio of logical amplitude error probability to phase error for rotation error is given as a function of spin $J$ for different value of phase error. 
Notably, this ratio exhibits an exponential trend, and for sufficiently large values of $J$, the logical amplitude error becomes negligible.
Consequently, there is no need for amplitude error correction in such cases To further illustrate the exponential suppression of the logical error arising from amplitude errors as a function of spin due to random rotation errors, in  \cref{fig:rot_error_b}, the ratio of logical amplitude error probability to phase error for rotation error is given as a function of spin $J$ for different value of phase error. 
Notably, this ratio exhibits an exponential trend, and for sufficiently large values of $J$, the logical amplitude error becomes negligible.
Consequently, there is no need for amplitude error correction in such cases, as {one does not need to pump the states back to $\{\ket{J}, \ket{-J}\}$ manifold as all the designed gates operate similarly in all the other lower kitten manifolds.}

\section{Photon scattering and optical pumping}
\label{sec:Ratio_optical_pumping_errors}
\begin{figure}[!ht]
\centering
\includegraphics[width=0.8\columnwidth]{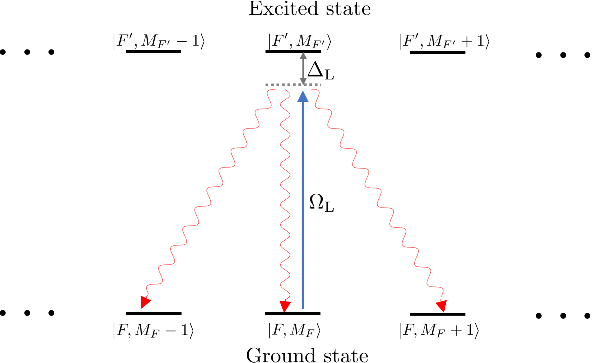}
\caption{The error process corresponding to the photon scattering and optical pumping for encoding a qudit in an atomic spin $\mathbf{F}$. The information is stored in the ground state and is controlled by laser light with Rabi frequency $\Omega_{\mathrm{L}}$ and detuning $\Delta_{\mathrm{L}}$ from an excited state manifold, with spin $\mathbf{F}'$. Absorption of a laser photon (here $\pi$-polarized) is followed by a spontaneous emission given by wavy lines.  The process causes amplitude errors and can collapse a cat-state to a single magnetic sublevel. }
\label{fig:decoherence_origin}
\end{figure}
Another major source of decoherence for the qubit encoded in a spin is the optical pumping arising from photon scattering when the spin are manipulated by laser light.
We consider here optical pumping arising from laser excitation with Rabi frequency $\Omega_{\mathrm{L}}$ and detuning $\Delta_{\mathrm{L}}$  from a dominant resonance. Absorption of a laser with polarization $\vec{\epsilon}_L$ is followed by a spontaneous emission of photon $\mathbf{e}_q$. A schematic of the error process corresponding to the photon scattering and optical pumping for atomic spins is shown in \cref{fig:decoherence_origin} for the case $\vec{\epsilon}_L = \mathbf{e}_0$. 

In this section, the spin angular momentum in which we encode the qudit is $\mathbf{F}$, and  $\mathbf{J}$ is the total angular momentum of the electrons. 
The jump operators for the optical pumping followed by photon scattering are, \cite{deutsch2010quantum}:
\begin{equation}
W_q=\sum_{F'}\frac{\Omega_{\mathrm{L}}/2}{\Delta_{FF'}+i\Gamma/2}(\bm{e}_q^{*}.\bm{D}_{FF'})(\vec{\epsilon}_L.\bm{D}_{FF'}^{\dagger}),
\label{eq:jump_operators_opt}
\end{equation} 
where $\Omega_{\mathrm{L}}$ is the Rabi frequency and $\Delta_{FF'}$ is the detuning between the ground state and excited with total spin $F$ and $F'$ respectively. 
$\Gamma$ is the characteristic linewidth of the excited state, $\vec{\epsilon}_L$ is the polarization of the laser, and $q={-1,0,1}$  represent the polarization of the scattered light.
$\bm{D}_{FF'}$ are the dimensionless raising operators from a ground state with total spin $F$ to an excited state with spin  $F'$ and see \cite{deutsch2010quantum} for a detailed analysis of these operators.

By decomposing the dyadic into irreducible tensors, one can derive a basis independent representation for the jump operators~\cite{deutsch2010quantum},
\begin{equation}
\begin{aligned}
&= \bm{e}_q^{*}.(\bm{D}_{FF'}\bm{D}_{FF'}^{\dagger}).\vec{\epsilon}_L\\
&=C_{J'FF'}^{0}\bm{e}_q^{*}.\vec{\epsilon}_L+iC^{1}_{J'FF'}(\bm{e}_q^{*}\cross\vec{\epsilon}_L).\bm{F}\\
&+C_{J'FF'}^{2}\left[\frac{(\bm{e}_q^*.\bm{F})(\vec{\epsilon}_L.\bm{F})+(\vec{\epsilon}_L.\bm{F})(\bm{e}_q^{*}.\bm{F})}{2}-\frac{1}{3}|\bm{e}_q^{*}.\vec{\epsilon}_L|\bm{F}^2\right]
\label{eq:jump_operators_optical_pumping_basis_independent}
\end{aligned}
\end{equation}
 where $J$ is the electron angular momentum. The above expression involves only angular momentum operators of the form $\bm{F}$ (rank-1) and $\bm{F}^2$ (rank-2), and thus for photon scattering and optical pumping the error operators are linear and quadratic powers of angular momentum operators.
Then the Lindblad master equation gives us: 
\begin{equation}
\begin{aligned}
    \frac{d\rho(t)}{dt} &=-i \left(H_\mathrm{eff}\rho(t)-\rho(t) H_\mathrm{eff}^\dag \right)+ \Gamma\sum_{i}W_q\rho(t) W_{q}^{\dagger}  \nonumber \\ 
&\equiv \mathcal{L}\rho(t).
\end{aligned}
\label{eq:evolution_of_the_density_matrix}
\end{equation}
where $\mathcal{L}$ is the Lindbladian and $H_\text{eff}=H-i\sum_q W_q^{\dagger}W_q/2$.

From the jump operators, one can find the probability of phase errors and amplitude errors by finding the overlap of the jump operators with the basis operators as given in \cref{eq:basis}.


\section{Correctable set of errors}
\label{sec:KL_conditions}
In this section, we find the set of correctable errors for the logical level encoding $\mathcal{C}_1$ in \cref{eq:concat_spin_cat}.
To find the correctable set of errors $\{E_a\}$, one can use the Knill-Laflamme conditions \cite{PhysRevA.55.900}:
\begin{equation}
\bra{\psi_i}E_a^{\dagger} E_b\ket{\psi_j}=C_{ab}\delta_{ij},
\label{eq:Knill-Laflamme conditions}
\end{equation}
where $i,j=\{0,1\}$ represents the codespace of interest.

The local angular momentum errors of interest here are of the form $J_x^lJ_y^mJ_z^n$. 
From the locality assumption of the errors, one can find that for the spin-cat encoding in \cref{eq:concat_spin_cat},
\begin{equation}
\bra{\psi_i}E_a^{\dagger} E_b\ket{\psi_j}=0 \hspace{0.2cm}\forall i\neq j.
\end{equation}
The next condition we need to satisfy for the spin-cat encoding is,
\begin{equation}
\bra{+_\mathrm{L}}E_a^{\dagger}E_b\ket{+_\mathrm{L}}=\bra{-_\mathrm{L}}E_a^{\dagger}E_b\ket{-_\mathrm{L}}
\label{eq:kl_diagonal}.
\end{equation}
From the locality assumption of the noise, this condition translates into two cases. 
In the first case the error operators $E_a$ and $E_b$ act on the same physical system, thus for the angular momentum errors the error correction condition in \cref{eq:kl_diagonal} becomes, 
\begin{equation}
    \begin{aligned}
        \bra{+}J_x^lJ_y^mJ_z^n J_x^{l'}J_y^{m'}J_z^{n'} \ket{+}=\bra{-}J_x^lJ_y^mJ_z^n J_x^{l'}J_y^{m'}J_z^{n'} \ket{-}.       
    \end{aligned}
    \label{eq:kl_local_same}
\end{equation}
Using an alternate definition of the spin-cat codes,
\begin{equation}
\begin{aligned}
\ket{\pm }=\frac{\mathds{1}\pm\exp(i\pi J_y)}{\sqrt{2}}\ket{J,-J},
\end{aligned}
\label{eq:basis_states}
\end{equation}
\cref{eq:kl_local_same} transforms into a compact expression:
\begin{equation}
\begin{aligned}
    &\bra{J,J } J_x^lJ_y^mJ_z^n J_x^{l'}J_y^{m'}J_z^{n'}\ket{J,-J}\\
    &=\bra{J,-J } J_x^lJ_y^mJ_z^n J_x^{l'}J_y^{m'}J_z^{n'}\ket{J,J}=0.
\end{aligned}
\label{eq:error_condition_3}
\end{equation}
Plugging the ladder operators,
\begin{equation}
\begin{aligned}
J_{+}&=J_x+i J_y\\
J_{-}&=J_x-i J_y
\end{aligned}
\label{eq:ladder_operators}
\end{equation}
into \cref{eq:error_condition_3}, and using the condition that  one needs at least $2J-1$ operations of $J_{+}$ or $J_{-}$ to make the overlap between the states $\ket{J,J}$ and $\ket{J,-J}$ non-zero, the error correction condition in \cref{eq:kl_local_same} simplifies to,
\begin{equation}
    l+m+n+l'+m'+n' \leq 2J-1.
\end{equation}
Thus we can correct the errors of the form $J_x^lJ_y^mJ_z^n$ if 
\begin{equation}
    l+m+n\leq \lfloor\frac{2J-1}{2}\rfloor.
\end{equation}

The second case for \cref{eq:kl_diagonal} is when the two error operators $E_a$ and $E_b$ act on different physical systems. For the angular momentum errors this simplifies to,
\begin{equation}
    \begin{aligned}
        &\bra{+}J_x^lJ_y^mJ_z^n\ket{+} \bra{+}J_x^{l'}J_y^{m'}J_z^{n'} \ket{+}\\
        &=\bra{-}J_x^lJ_y^mJ_z^n\ket{-} \bra{-}J_x^{l'}J_y^{m'}J_z^{n'} \ket{-}.
    \end{aligned}
\end{equation}

\noindent Again using the \cref{eq:basis_states} and \cref{eq:ladder_operators}, the error correction condition is given as:
\begin{equation}
    \begin{aligned}
        l+m+n &\leq 2J-1,\\
         l'+m'+n' &\leq 2J-1.
    \end{aligned}
\end{equation}

\noindent Hence the  spin-cat encoding can correct all the errors of the form,
\begin{equation}
\begin{aligned}
\mathcal{E}_K=
\left\{
J_x^{l}J_y^{m}J_z^{n}; 0\leq l+m+n\leq K=\lfloor\frac{2J-1}{2} \rfloor\right\}.
\end{aligned}
\end{equation}

\section{Action of the SU(2) operators}
\label{sec:coefficients_rotation}
The Euler angle representation of an $\mathrm{SU}(2)$ operator  $V=\exp(-i \theta \hat{n}.\mathbf{J})$ is,
\begin{equation}
    V(\alpha,\beta,\gamma)=\exp(-i \theta \hat{n}.\mathbf{J})= e^{-i\alpha J_z} e^{-i\beta J_y} e^{-i\gamma J_z}.
\end{equation}
The Wigner $D$ matrix defined in \cref{eq:Wigner_D_matrix} can be expressed in terms of Euler angles as,
\begin{equation}
    \begin{aligned}
        D_{q,q'}(\alpha,\beta,\gamma)&=\bra{k,J_z=q'} \exp(-i \theta \hat{n}.\mathbf{J}) \ket{k,J_z=q}\\
        &=e^{-iq'\alpha} d_{q,q'}(\beta) e^{-iq \gamma}.
    \end{aligned}
\end{equation}
Hence, deriving from the definitions of the spherical tensor operators in \cref{eq:spherical_tensor_operators}, the operators in \cref{eq:basis}, and the inherent properties of the Wigner $d$ matrices,
\begin{equation}
    d_{q, q'}=(-1)^{q-q'}d_{-q,-q'},
\end{equation}
we find the action of an SU(2) rotation acting on the error operator, Eq. (\ref{eq:basis}) is
\begin{equation}
    \begin{aligned}
       & V S^{(k)}_q V^{\dagger}\\
       &= \sum_{q'}\frac{f_{q,q'}(\vec{\theta})}{\sqrt{2}} \left(T^{(k)}_{q'}+(-1)^{q-q'+k}e^{-2i\left(q\alpha+q'\beta\right)}T^{(k)}_{-q'}\right),\\
       &=\sum_{q'} f_{q,q'} S^{(k)}_q+ \frac{\widetilde{f}_{q,q}}{2} \left({F}^{(k)}_q-A^{(k)}_q\right).
    \end{aligned}
\end{equation}
where to lighten the notation we defined,
\begin{equation}
    \widetilde{f}_{q,q}=(-1)^k \left[1-(-1)^{q-q'}e^{-2i\left(q\alpha+q'\beta\right)}\right] f_{q,q'}.
\end{equation}
Thus,
\begin{equation}
    \begin{aligned}
        V S^{(k)}_q V^{\dagger}&=\sum_{q'} g_{q,q'} S^{(k)}_q+ \widetilde{g}_{q,q}A^{(k)}_q
    \end{aligned}
\end{equation}
where we have defined,
\begin{equation}
    \begin{aligned}
         g_{q,q'}&= f_{q,q'}+\frac{\widetilde{f}_{q,q'}}{2},\\
           \widetilde{g}_{q,q'}&=-\frac{\widetilde{f}_{q,q'}}{2}.
    \end{aligned}
\end{equation}
Similarly,
\begin{equation}
    \begin{aligned}
        V A^{(k)}_q V^{\dagger}&=\sum_{q'} h_{q,q'} S^{(k)}_q+ \widetilde{h}_{q,q}A^{(k)}_q
    \end{aligned}
\end{equation}
where again for simplification of notation,
\begin{equation}
    \begin{aligned}
         h_{q,q'}&= \frac{(-1)^k \left[1+(-1)^{q-q'}e^{-2i\left(q\alpha+q'\beta\right)}\right] f_{q,q'}}{2},\\
           \widetilde{h}_{q,q'}&=f_{q,q'}-h_{q,q'}.
    \end{aligned}
\end{equation}
Thus the action of the $\mathrm{SU}(2)$ does not change the rank of the error operators, $ A^{(k)}_q, {F}^{(k)}_q$ and obey the condition given in  \cref{eq:fault_tolerant_condition}.

\section{Rotating the ground and excited manifold differently using optimal control}
\label{sec:Rotating_the_ground_and_excited_manifold}
To implement the rank-preserving CNOT gate in \cref{fig:fig_cnot}, one needs to implement $X=\exp(-i\pi J_x)$ gate on the auxiliary manifold while applying the identity operator on the Rydberg manifold.  
For the specific choice of auxiliary and Rydberg states considered, we have the Hamiltonian in the rotating field as given by \cref{eq:Hamiltonians}.
As we are dealing with $\mathrm{SU}(2)$ representations of the spin $J$, the problem is isomorphic to the simultaneous control of two two-level systems/two qubits with different Rabi frequencies and different detuning. 
The objective would be to apply a Pauli $X$ operation on the first qubit and identity on the second system.
This problem has a Quantum-Speed-Limit(QSL) of $\pi/\Omega_{\mathrm{rf}}$\cite{Qcontrol_QSL_Hegerfeldt}.

\begin{figure}[!ht]
    \centering
   \includegraphics[width =1\columnwidth]{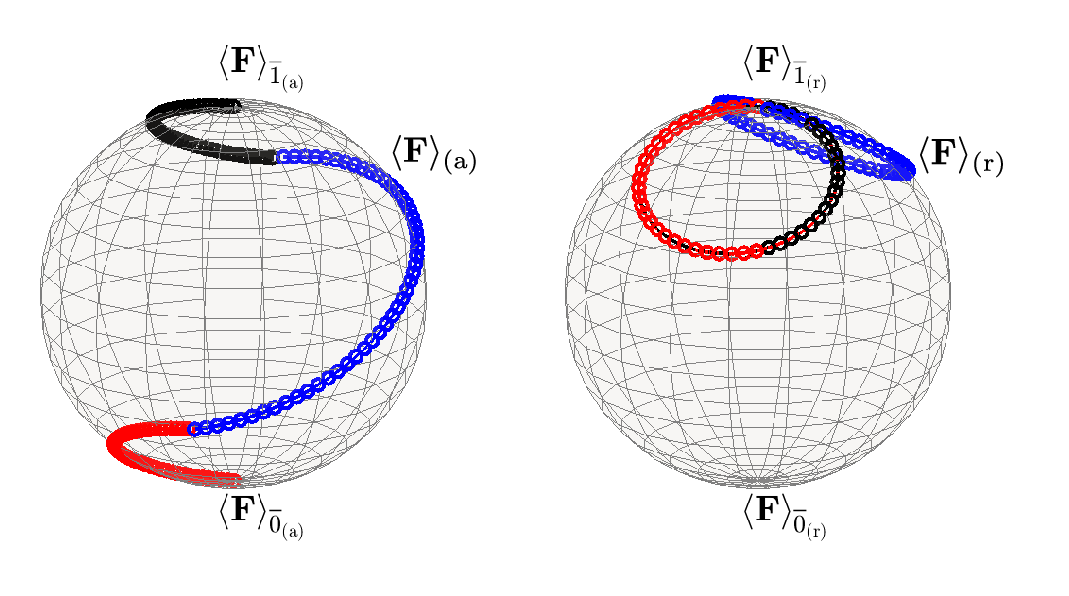} 
    \caption{ Evolutions of the spin vector $\langle \vec{F} \rangle$ for the auxiliary (a) and Rydberg (r) manifolds resulting from rf-driven Larmor precession with time-varying phases in \cref{eq:Hamiltonians} for piecewise constant function with $3$ time steps with a total time  $T_{\mathrm{tot}}=3\pi/\Omega_{\mathrm{rf}}$ and $\omega_0=5 \Omega_{\mathrm{rf}}$. 
    For the specific choice of parameters, an $X$ gate acts on the auxiliary manifold and transfers the population from $\overline{0}_{\mathrm{a}}$-subspace to $\overline{1}_{\mathrm{a}}$-subspace and vice-versa.
However, for the  Rydberg manifold, the pulse sequence acts as an identity operator, and the population in the  $\overline{0}_{\mathrm{r}}$ and  $\overline{1}_{\mathrm{r}}$ subspaces remain unaffected.}
    \label{fig:bloch_3_variable}
\end{figure}

Since $\Omega_{\mathrm{r}}=2\Omega_{\mathrm{a}}$, a pulse of length $\pi/\Omega_{\mathrm{a}}$ would cause a full Rabi rotation in the Rydberg manifold and only a half rotation in the auxiliary manifold. 
By choosing the phases of {N such} pulses, $\vec{\phi}$ in \cref{eq:Hamiltonians}, 
 one can use quantum optimal control algorithms to implement the desired transformation.
The minimum number of pulses $N$ required depends on the ratio $\omega_0/\Omega_{\mathrm{rf}}$. 
While a solution with $N=2$ only exists when $\omega_0/\Omega_{\mathrm{rf}}=3$, a solution with $N=3$ is possible if $\omega_0/\Omega_{\mathrm{rf}}<3\sqrt{3}$ for example the case of $\omega_0=5\Omega_{\mathrm{rf}}$ and $T=3\pi/\Omega_{\mathrm{rf}}$ is given in \cref{fig:bloch_3_variable}.
The overall trend is that with an increasing ratio $\omega_0/\Omega_{\mathrm{a}}$, we need a larger $N$. This protocol is similar to \cite{Levine_Pichler_gate}, and takes $N\pi/\Omega_{\mathrm{a}}$, which is longer than the QSL. We can use waveforms with a large number of steps to implement a gate in the minimum time $\pi/\Omega_{\mathrm{rf}}$, as shown in the example below.
\vspace{0.2cm}

\begin{figure}
   \centering
\includegraphics[width=\columnwidth]{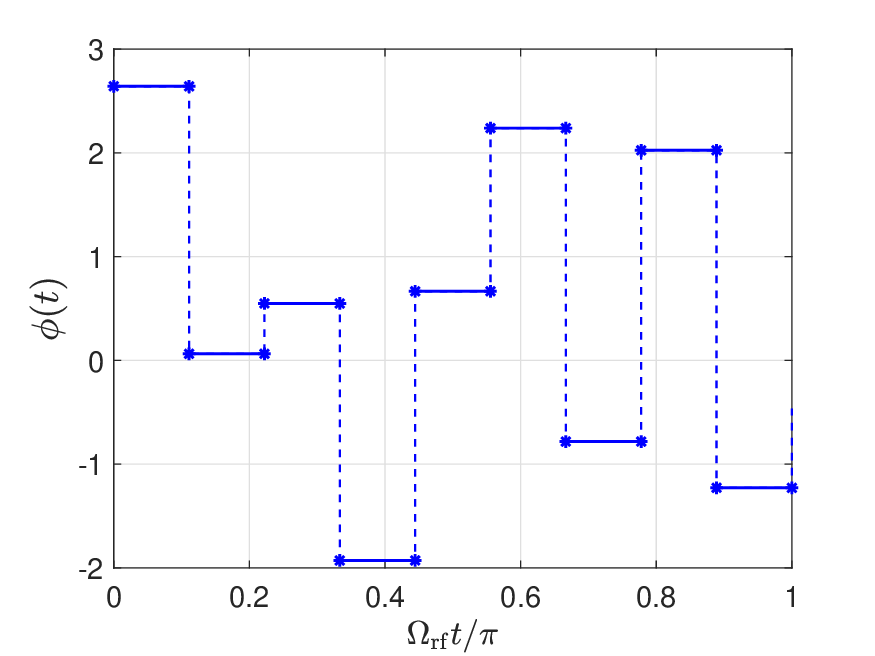}
\caption{ The phase $\phi(t)$ which generates an $R=\exp(i\pi J_z)$ for the auxiliary manifold and an identity in the Rydberg manifold, which can be used to implement the rank-preserving $\mathrm{CZ}$ gate.
The total time is $\Omega_{\mathrm{rf}}T=\pi$, which is divided into $10$ equal time steps with  $\omega_0=3\Omega_{\mathrm{rf}}$ and pulse sequence is found using the quantum optimal control {algorithm} GRAPE.}
\label{fig:zz_pulse}
\end{figure}

 Using the  Hamiltonians in \cref{eq:Hamiltonians}, one can also optimize the phase $\phi$ to implement a gate $R(\theta)=\exp(-i\theta \hat{\bm{n}}.\mathbf{J})$ in the auxiliary manifold and identity on the Rydberg manifold. 
For example, the pulse scheme for the $R=\exp(i\pi J_z)$ for the auxiliary manifold, which can be used to implement the rank-preserving $\mathrm{CZ}$ gate is given in \cref{fig:zz_pulse}.
The total time is $\Omega_{\mathrm{rf}}T=\pi$ and total time is divided into $10$ equal time steps with  $\omega_0=3\Omega_{\mathrm{rf}}$.

Finally, for $\omega_0\gg\Omega_{\mathrm{rf}}$, a field that is resonant for the auxiliary spin will be far off-resonant for the Rydberg manifold. So we can implement any desired transformation $\mathrm{SU}(2)$ operation in the auxiliary subspace without disturbing the Rydberg manifold populations.

 \section{Implementing Hadamard gate from the Physical level gates}
 \label{sec:Hadamard_gate}
 \begin{figure}
    \centering
    \includegraphics[width=0.8\columnwidth]{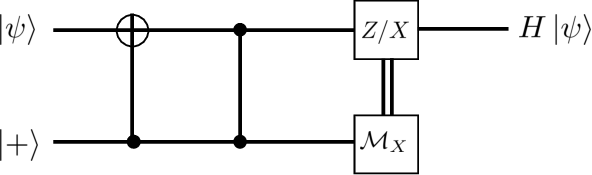}
    \caption{Circuit implementing a fault-tolerant Hadamard gate using the physical level gates for the spin-cat encoding.
    This differs from the standard implementation as we use both CNOT and CZ gate to implement the action of the target unitary of interest.}
    \label{fig:hadamard_gate}
\end{figure}
The  physical level gates for the spin-cat encoding are given as,
 \begin{equation}
    \{\mathcal{M}_z,\mathcal{M}_X,\mathcal{P}_{\ket{+}},\mathcal{P}_{\ket{0}},\mathrm{CNOT},X,Y,Z,ZZ(\theta)\}.  
 \end{equation}
 The Hadamard gate is not in the universal gate set as it does not preserve the rank. Here we show the implementation of the Hadamard gate using the rank-preserving physical level gates and an ancilla qubit.
 The circuit diagram corresponding to a teleportation-based scheme for the Hadamard gate is given in the \cref{fig:hadamard_gate}. 
Consider an initial arbitrary state,
 \begin{equation}
     \ket{\psi}=\alpha \ket{0}_k+\beta\ket{1}_k,
 \end{equation}
 and ancilla state,
 \begin{equation}
     \ket{+}_0=\frac{1}{\sqrt{2}}\left(\ket{0}+\ket{1}\right).
 \end{equation}
 Define $\ket{\phi}=\ket{\psi}\otimes \ket{+}_0$, then
 \begin{equation}
 \begin{aligned}
      &\mathrm{CNOT}\ket{\phi}= \mathrm{CNOT}\ket{\psi}\otimes \ket{+}_0\\
      &=\frac{1}{\sqrt{2}}\left(\alpha \ket{0}_k\ket{0}+\alpha \ket{1}_k\ket{1}+\beta \ket{1}_k\ket{0}+\beta \ket{0}_k\ket{1}\right),
 \end{aligned}    
 \end{equation}
and
 \begin{equation}
 \begin{aligned}
    \mathrm{CZ}\text {  } \mathrm{CNOT}\ket{\phi}&=\left(\alpha \ket{-}_k+\beta\ket{+}_k\right)\ket{+}\\
     &+\left(\alpha \ket{+}_k-\beta\ket{-}_k\right)\ket{-}.
 \end{aligned}     
 \end{equation}
 Thus one can act $Z$ or $X$ gate depending on the measurement of the $X$ operator in the ancilla to get the state,
 \begin{equation}
     H\ket{\psi}=\alpha \ket{+}_k+\beta\ket{+}_k,
 \end{equation}
and implement the action of the Hadamard gate.

\section{Implementing the Logical operator}
\label{sec:implementing_the_logical_operator}

 \begin{figure*}[!ht]
    \centering
    \subfloat[]{\includegraphics[width =0.8\columnwidth]{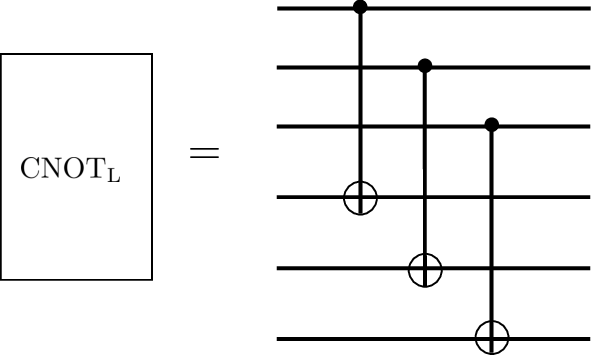} \label{fig:Fig_19_a}}\hspace*{1.9em}
    \subfloat[]{\includegraphics[width =1\columnwidth]{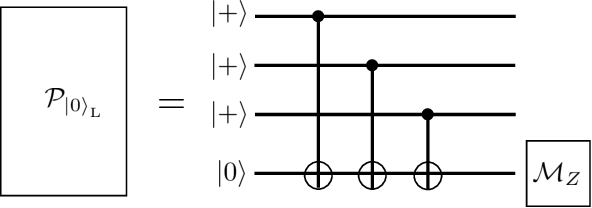}  \label{fig:Fig_19_b}}\\
    \subfloat[]{\includegraphics[width =.9\columnwidth]{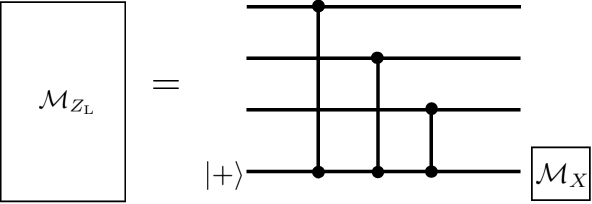} \label{fig:Fig_19_c}}
    \hspace*{1.9em}
    \subfloat[]{\includegraphics[width =0.9\columnwidth]{logical_mz.eps}  \label{fig:Fig_19_d}}
    \caption{Circuits implementing logical level gates in $\mathcal{C}_1$ using the physical level gates.
    (a)  {Logical $\mathrm{CNOT}$ $\mathrm{CNOT}_{\mathrm{L}}$. To implement $\mathrm{CNOT}_{\mathrm{L}}$, we apply physical $\mathrm{CNOT}$ gates transversally on all qubit pairs}.
(b) Preparation of $\ket{0}_{\mathrm{L}}$ $\mathcal{P}_{\ket{0}_{\mathrm{L}}}$. $\ket{0}_{\mathrm{L}}$ is prepared by initializing the system with the state $\mathcal{P}_{\ket{+}_{\mathrm{L}}}$ and measuring the parity.
To measure the parity we use an ancilla initialized with $\mathcal{P}_{\ket{0}}$ and use physical CNOT gates followed by measuring the $\mathcal{M}_Z$, the final state  is  $\ket{0}_{\mathrm{L}}$ or $\ket{1}_{\mathrm{L}}$ for the measurement outcomes $1$ and $-1$ respectively. 
(c) {The Logical $Z$ measurement $\mathcal{M}_{Z_{\mathrm{L}}}$}. An ancilla state is prepared in $\ket{+}$ and physical CZ gates {with the data qubits are applied} followed by measuring the ancilla in the $X$ basis.
(d) Logical $X$ measurement $\mathcal{M}_{X_\mathrm{L}}$. 
The logical $X$ is measured by applying the physical CNOT gates and then measurement along $X$.}
    \label{fig:logical_gate}
\end{figure*}
In this section, we demonstrate the universal gate set at the logical level with the physical level gates for the spin-cat encoding. 
The rank-preserving physical level gates for the spin-cat encoding are,
\begin{equation}
    \{\mathcal{M}_Z,\mathcal{M}_X,\mathcal{P}_{\ket{+}},\mathcal{P}_{\ket{0}},\mathrm{CNOT},X,Y,Z\}.
\end{equation}
Consider a universal gate set,
\begin{equation}
    \{P_{\ket{0}_\mathrm{L}},P_{\ket{+}_\mathrm{L}},\mathcal{M}_{X_\mathrm{L}},\mathcal{M}_{Z_\mathrm{L}},\mathrm{CNOT}_\mathrm{L}\} \cup\{P_{\ket{i}_\mathrm{L}},P_{\ket{T}_\mathrm{L}}\}.
    \label{eq:logical_level_gate_app}
\end{equation}
{In the above equation, the first set generates the Clifford operations and the second set generates the non-Clifford states to complete the universal gate set, and $\mathcal{P}$ refers to preparation and $\mathcal{M}$ denotes measurement.}
The  logical preparation of the $\mathcal{P}_{\ket{+}_\mathrm{L}}$ can be done transversally by preparing the $\mathcal{P}_{\ket{+}}$ in the individual systems.
For example in the case of three physical systems, the logical level state preparation is,
\begin{equation}
    \begin{aligned}
         \mathcal{P}_{\ket{+}_\text{L}}= \ket{+}_\text{L}&=\ket{+++}.
              \end{aligned}
\end{equation}
In a similar fashion, the construction of additional logical-level gates follows the approach detailed in \cite{Aliferis2008fault,puri2020bias}.
Comprehensive details for the implementation of all other logical gates are provided in \cref{fig:logical_gate}.
In (a), the $\mathrm{CNOT}_{\mathrm{L}}$ is implemented using the physical CNOT gates.
One can implement the $\mathrm{CNOT}_{\mathrm{L}}$ by transversal application of the CNOT gates.
In (b), the $\mathcal{P}_{\ket{0}_{\mathrm{L}}}$ is prepared by initializing the system with the state $\mathcal{P}_{\ket{+}_{\mathrm{L}}}$ and measuring the parity.
To measure the parity we use an ancilla initialized with $\mathcal{P}_{\ket{0}}$ and use physical CNOT gates followed by measuring the $\mathcal{M}_Z$, the final state  is  $\mathcal{P}_{\ket{0}_{\mathrm{L}}}$ and $\mathcal{P}_{\ket{1}_{\mathrm{L}}}$ for the measurement outcomes $1$ and $-1$ respectively. 
(c) implements the logical measurement of $Z$ with an ancilla state prepared in $\ket{+}$ and physical CZ gates followed by measuring the ancilla in the $X$ basis.
Finally (d) implements the logical measurement by applying the physical CNOT gates and measurement of $X$.
{Access to the gate $ZZ(\theta)$ allows one to construct the non-Clifford part of the universal gate set with high-fidelity as studied in detail in \cite{zzPoulin}.}
\section{Toffoli gate}
\label{sec:Toffoli_gate}

\begin{figure}
    \centering
    \includegraphics[width=\columnwidth]{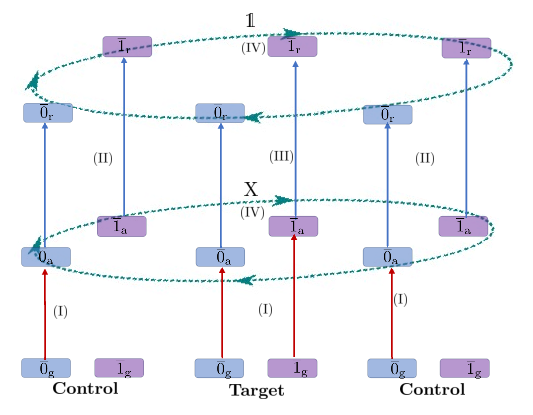}
    \caption{  Protocol for  a rank-preserving Toffoli {gate} for spin-cat encoding using $\mathrm{SU}(2)$ operations. 
 Similar to the rank-preserving CNOT gate \cref{fig:fig_cnot}, we implement the Toffoli gate in the ground state of $^{87}$Sr and the physical setting is the same as given in \cref{fig:Fig_2_a}.
We consider a geometry of atoms such that the nearest neighbors are constrained by the Rydberg blockade, but the next-nearest neighbors are not constrained. 
In step I the population is promoted to the auxiliary manifold in the atoms. 
{In} the control atoms we only promote the population of the $\overline{0}$-subspace whereas for the target atom, the population from both the $\overline{0}$ and $\overline{1}$ subspaces are promoted to the auxiliary state.
In step II, we transfer the population between the auxiliary and the Rydberg manifolds of the control atoms.
In step III, we transfer the population from the auxiliary to the Rydberg manifold of the target atom. However, due to the Rydberg blockade, this population transfer only happens when both the control atoms are in $\overline{0}$-subspaces. 
If even one of the control atoms is in $\overline{1}$-subspace this transition is blockaded. 
Then similar to the rank-preserving CNOT gate, in step IV we implement aa $X=\exp(-i\pi J_x)$ gate in the auxiliary manifold and an identity operator in the Rydberg manifold.   
Finally, we will transfer all the states back to the ground state by acting steps III-I in reverse, thus implementing a rank-preserving Toffoli gate for the spin-cat encoding.}
    \label{fig:fig_toffoli}
\end{figure}
One can generalize the rank-preserving CNOT gate in \cref{fig:fig_cnot} to construct a Toffoli gate, also known as a controlled-controlled NOT gate. 
\cref{fig:fig_toffoli} gives the protocol for creating the rank-preserving Toffoli gate for the spin-cat encoding using only $\mathrm{SU(2)}$ interactions. 
Again, similar to the rank-preserving CNOT gate, the Toffoli gate is implemented in the ground state of $^{87}$Sr.
The key to the scheme is the availability of special geometries for the neutral atoms \cite{Levine_Pichler_gate,Lukin_Nature_2022}. 
Here we use a geometry such that {for three linearly arranged atoms,} the nearest neighbors are constrained by the Rydberg blockade, but the next-nearest neighbors are not constrained by it.
{The central atom acts as the target atom while its two neighbors are the control atoms.}

In step I of the Toffoli gate, the population is promoted to the auxiliary state. For the case of the control atoms we only promote the population of the $\overline{0}$-subspace whereas for the target atom, the population from both the $\overline{0}$ and $\overline{1}$ subspaces are promoted to the auxiliary state.
In step II, we use a pulse sequence similar to the \cref{fig:Fig_waveform_c} to transfer the population between the auxiliary and the Rydberg state of the control atoms using $\pi$ polarized light.
In step III we apply the same pulse sequence as in step II to the target atom however, due to the Rydberg blockade, the population transfer between the auxiliary and Rydberg state only happens when both the control atoms are in $\overline{1}$-subspace. 
Then similar to the case of the rank-preserving CNOT gate in step IV, we implement a $X=\exp(-i\pi J_x)$ gate in the auxiliary manifold and an identity operator in the Rydberg manifold.  
Finally, we will transfer all the states back to the ground state by acting steps III-I {in reverse,} thus implementing a rank-preserving Toffoli gate for the spin-cat encoding up to local rotations.

Thus when one of the control atoms is in the $\overline{0}$-subspace, $X$ gate is applied target atom, and when both the control atoms are in the $\overline{1}$-subspace, the target atoms remain unchanged.
This is the Toffoli gate up to a local {$X=\exp(-i\pi J_x)$} rotation on the target atom. 
{In a realistic implementation of this protocol, one needs to consider the finite range of the Rydberg blockade while implementing the pulses for the Toffoli gate. To address this challenge, the effect of finite blockade effect can be addressed by using an appropriate control Hamiltonian in the quantum optimal control and finding control pulses that mitigate the undesirable effect of the finite blockade. }
\section{Alternate approaches for cat-state preparation and measurement of $X$ }
One can use alternative approaches than quantum optimal control for cat-state preparation and measurement of $X$. In this section, we detail some of those approaches. 
For example, one can use an adiabatic approach and one-axis twisting to create a spin-cat state.

I) Adiabatic approach. Starting with an initial state $\ket{J,J_z=J}$ and evolving the Hamiltonian 
\begin{equation}
    H(s)=(1-s)J_x-\frac{s}{2J}J_z^2,
\end{equation}
adiabatically {from $s=0$ to $s=1$} guarantees the final state to be close to a cat state $\ket{+}$ \cite{puri2017engineering}. 
This can be implemented in atomic systems using a combination of tensor light shifts and rf rotation~\cite{Paul_experiment_Cs_2007,omanakuttan2021quantum}.

 II) One-axis twisting: Using a time-independent Hamiltonian, $H=\beta J_z^2$, for a certain time $T=\pi/(2\beta)$, one can evolve a spin coherent state along $J_x$ to prepare a high-fidelity cat state.
\begin{equation}
    \ket{+}=\exp(-i\pi J_x)\exp(-i\frac{\pi}{2}J_z^2)\ket{J,J_x=J}.
\end{equation}

Including the effect of decoherence due to photon scattering and optical pumping for  $^{87}$Sr, we find the fidelity for one-axis twisting is $0.9998$ whereas for the adiabatic preparation, one can achieve a fidelity of $0.9889$.

Similarly one can use an alternative approach to measure $X$, in particular, to know if the ancilla state is in $\ket{+}$ or $\ket{-}$. 
We can adiabatically rotate the states using the Hamiltonian
\begin{equation}
    H(s)=-(1-s)J_z^2/(J)+s J_x,
\end{equation}
which implements the following transformations:
\begin{equation}
    \begin{aligned}
        \ket{+}_0&\to \ket{J,J_x=J},\\
        \ket{-}_0 &\to \ket{J,J_x=J-1},
    \end{aligned}
    \label{eq:M_x_transformation_adiabatic}
\end{equation}
and then then measuring  $J_x$.

To evaluate the accuracy of $X$ measurement, we define the target isometry as: 
\begin{equation}    V_{\mathrm{targ}}=\ket{J,J_x=J}\bra{+}+\ket{J,J_x=J-1}\bra{-}.
\end{equation}
The implemented isometry using the adiabatic approach is given as,
\begin{equation}
    V= e^{-\int \mathcal{L}(s)ds} V(0)
\end{equation}
where $\mathcal{L}(s)$ is the Lindbladian including the effects of decoherence and 
\begin{equation}
    V(0)=\ket{+}\bra{+}+\ket{-}\bra{-}.
\end{equation}
Thus the fidelity for the implementation of the isometry is defined as:
\begin{equation}
    \mathcal{F}_{\mathrm{iso}}=\frac{1}{4}\lvert\Tr(V_{\mathrm{targ}}V^{\dagger})\rvert^2.
    \label{eq:accuray_equation_app}
\end{equation}
This approach is similar to the approach taken in bosonic cat qubits \cite{puri2019stabilized}. To measure $J_x$, we first implement the unitary transformation $U=\exp(-i\pi/2 J_y)$ to rotate the basis to $\ket{J,J_z}$ and then perform the readily accessible measurement $\mathcal{M}_Z$ which we can in principle achieve with a fidelity larger than $99\%$ \cite{barnes2022assembly}. 
Including the effects of optical pumping as discussed in \cref{sec:Ratio_optical_pumping_errors}, one can implement this transformation with a fidelity of $\mathcal{F}_{\mathrm{iso}}=0.98$ for the $^{ 87}$Sr nuclear spin qudit.

\section{Error correction without measurement }
\label{sec:measurement_free_ec}

An alternative to syndrome-based quantum error correction is measurement-free quantum error correction (MFQEC) \cite{PhysRevLett.117.130503,PhysRevA.97.012318,li2011recovery,premakumar2020measurement}.
The standard syndrome-based error correction is given by recovery operation:
\begin{equation}
     R(\rho)=\sum_i U_i M_i \rho M_i^{\dagger} U_i^{\dagger},
\end{equation}
where for a general state $\rho$, $M_i$ is the syndrome measurement and $U_i$ is the correction unitary according to the outcome of the syndrome measurement. 

MFQEC is based on the unitary operator $V$, which couples the data and ancilla qubits.
The action of which is given as,
\begin{equation}
    V\ket{\psi} \ket{0}=\sum_i \left(U_i M_i \otimes \mathds{1} \right)\ket{\psi} \ket{i}.
\end{equation}
Defining $\rho=\sum_{kl}\alpha_{kl}\ket{\psi}_k\bra{\psi}_l$, we can find that,
\begin{equation}
    V\rho \otimes \ketbra{0}{0} V^{\dagger}=\sum_{k,l,i,j}\alpha_{kl}U_iM_i\ket{\psi}_k\bra{\psi}_l M_j^{\dagger} U_j^{\dagger} \otimes \ketbra{i}{j}
\end{equation}
 Partial tracing of the ancilla gives,
\begin{equation}
    \rho_{\mathrm{rec}}=\sum_{i}U_i M_i \rho M_i^{\dagger} U_i^{\dagger}.
\end{equation}
Thus the MFQEC is equivalent to syndrome-based error correction and the key for MFQEC is a specific unitary gate between the ancilla and the data.

One can consider a fault-tolerant  MFQEC scheme for the amplitude errors.
The syndrome for the amplitude errors is the eigenvalue of $J_z^2$, which can be extracted by the projective measurement, 
\begin{equation}
    M_k=\ket{+}_k\bra{+}_k+\ket{-}_k\bra{-}_k,
\end{equation}
where $0\leq k\leq (2J-1)/2$.
Recovery unitaries corresponding to  the projective measurement outcomes are
\begin{equation}  
\begin{aligned}    U_k&=\ket{+}_0\bra{+}_k+\ket{+}_k\bra{+}_0+\ket{-}_0\bra{-}_k+\ket{-}_k\bra{-}_0\\
&+\sum_{j\neq k, j\neq 0} \ket{+}_j\bra{+}_j+\ket{-}_j\bra{-}_j,
\end{aligned}
\end{equation}
which takes the state from the subspace,
\begin{equation}
    \{\ket{+}_k, \ket{-}_k\} \to \{\ket{+}_0, \ket{-}_0\}.
\end{equation}
Consider the following unitary operator, using the definitions from \cref{eq:projectors,eq:logical_paulis,eq:cnot_def} the product of three alternating CNOT gates can be written as:
\begin{equation}
\begin{aligned}
     V_s&=\Pi_{\overline{0}}\otimes \Pi_{\overline{0}}+\Pi_{\overline{1}}\otimes \Pi_{\overline{1}}\\
     &+X\Pi_{\overline{0}} \otimes X\Pi_{\overline{1}} +X\Pi_{\overline{1}}\otimes X\Pi_{\overline{0}},
\end{aligned}   
\label{eq:V_s_gate}
\end{equation}
Consider the following states, 
\begin{equation}
    \begin{aligned}
        \ket{\psi}_k&=\alpha \ket{+}_k+\beta\ket{-}_k,\\
        \ket{\phi}_l&=\gamma \ket{+}_l+\delta\ket{-}_l,\\
    \end{aligned}
\end{equation}
where $\alpha,\beta, \gamma, \text{ and } \delta$ are arbitrary complex amplitudes. 
The action of the $V_s$ on the state, $\ket{\xi}=\ket{\psi}_k\otimes \ket{\phi}_l$ gives,
\begin{equation}
    V_s \ket{\xi}=\ket{\phi}_k\otimes \ket{\psi}_l.
    \end{equation}
 Thus $V_s$ gate swaps the information between two kitten or cat states.
 The circuit diagram for the $V_s$  gate for a qubit encoded in the qudit is given in \cref{fig:Fig_swap_a}.

When the second qudit is prepared in  $\ket{+}_0$ state, as shown in \cref{fig:Fig_swap_b}, 
the application of the $V_s$ gate gives
\begin{equation}
   \ket{\phi}= V_s\ket{\psi}\ket{+}_0=\ket{+}_k \otimes \left(\alpha\ket{+}_0+\beta \ket{-}_0\right). 
\end{equation}
The above state can also be written as,
\begin{equation}
    \ket{\phi}=\sum_k U_k M_k \ket{\psi} \ket{+}_k,
\end{equation}
where the notion of data and ancilla qubits are swapped for convenience. 
Thus the unitary operator $V_s$ followed by partial tracing implements the desired recovery operation.
Thus one can correct the amplitude error fault tolerantly using a combination of two rank-preserving CNOT gates and fresh $\ket{+}_0$ state.

For fault-tolerant gadgets, one needs to repeat the phase and amplitude error correction multiple times and one needs to ensure that these two error correction steps commute with each other.
The  phase error correction \cref{fig:circuit_phase_error_correction}  commutes with measurement-free error correction of the amplitude error and the details of the calculation are given in 
\cref{sec:commutativity_of_error_operators}.

\subsection{Upper bounds on the probability of
the logical error in the amplitude error correction}
\label{sec:upper_bounds_amplitude}
In this section, we provide a detailed analysis to find an upper bound on the probability of a logical error in the amplitude error correction used in the error-corrected logical CNOT gadget in \cref{fig:fault_tolerant_cx_gadget}.

First, consider the case where ancilla is prepared perfectly, i.e., we have $\rho_A=\ket{+}_0$ and $p_i=0$ for $i\neq 0$ in \cref{fig:swapping_animation}. In this case, a logical amplitude error occurs after $s$ faulty CNOT gates if they create at least $k_{\mathrm{max}}=\lfloor (2J+1)/2\rfloor$ many jumps, the probability of which we denote by $q(s,k_{\mathrm{max}})$. The number of CNOT gates $s$ is determined by the number of phase error corrections that appear before an amplitude correction, in addition to the two CNOT acting in the amplitude error correction itself. To find the probability $q(s,k_{\mathrm{max}})$, we note that each physical CNOT gate can create one or two jumps with probabilities $p_1$ and $p_2$ respectively, and therefore we need to add the probabilities of cascades of one and two jumps that can create more than $k_{\mathrm{max}}$ jumps. Therefore $q(s,k_{\mathrm{max}})$ can be written as
\begin{equation}
    q(s,k_{\mathrm{max}})=\sum_{i}\lambda_i(s,k_{\mathrm{max}}),
    \label{eq:amplitude_error_ideal}
\end{equation}
where $\lambda_i$ represents the probability of one path such that we have at least $k_{\mathrm{max}}$ jumps. For example, consider the case of $s=4$ and $J=9/2$, then $\lambda_i$ represents all the possible combinations of one and two jumps, such that the total sum of these jumps is at least $5$. 
One such possibility is a combination of $(1,1,1,2)$ where we have one jump occurring at the first three CNOTs and two jumps occurring at the last CNOT.



When the ancilla is imperfect, for example, if it is prepared in $\ket{+}_k$ state rather than $\ket{+}_0$, one needs to find the paths that create $k_{\mathrm{max}}-k$ many jumps. Thus we get, 
\begin{equation}
q(s,k_{\mathrm{max}}|k)=\sum_i \lambda_i(s,k_\mathrm{max}|k),
\end{equation}
where $\lambda_i(s,k_{\mathrm{max}}\lvert k)$ is the probability of a path where we have at least $k_{\mathrm{max}}$ jumps given that we already had $k$ jumps to start with.

We repeat the amplitude error correction $r_2$ many times in one error-corrected logical CNOT gate. Thus the upper bound of the logical amplitude error probability after $r_2$ rounds of error correction in \cref{fig:fault_tolerant_cx_gadget} is,
\begin{equation}
\epsilon^{\mathrm{amp}}=r_2\left(\sum_{k=0}^{4}q(s,k_{\mathrm{max}}|k)p_k\right).
\end{equation}
where $p_k$ is the probability of ancilla starting at $\ket{+}_k$.

\vspace{0.7cm}
\section{Commutativity of the Error correction steps}
\label{sec:commutativity_of_error_operators}

\begin{widetext}
    The error correction for the spin-cat encoding follows two steps. 
The first step is the phase error correction in \cref{fig:circuit_phase_error_correction} and the second step is the measurement-free error correction for correcting amplitude errors given in \cref{fig:SWAP_gate_qubit_1}.
For fault-tolerant gadgets, one needs to repeat these steps multiple times and we need to ensure that these two error correction steps commute with each other such that the errors do not proliferate uncontrollably.
For this, we  need to satisfy,
\begin{equation}
    \mathcal{R}_{\mathrm{amp}} \mathcal{R}_{\mathrm{ph}}\left(\mathcal{E}\left(\rho\right)\right)= \mathcal{R}_{\mathrm{ph}} \mathcal{R}_{\mathrm{amp}}\left(\mathcal{E}\left(\rho\right)\right),
\end{equation}
where $\mathcal{R}_{\mathrm{amp}}, \mathcal{R}_{\mathrm{ph}}$ are the recovery maps corresponding to the amplitude and phase error correction respectively.
The recovery map for the amplitude error can be expressed in terms of the Kraus operators as,
\begin{equation}
    \mathcal{R}_{\mathrm{amp}}(\rho)=\sum_{j,i} M_{j,i}^{\mathrm{amp}} \rho \left(M_{j,i}^{\mathrm{amp}}\right)^{\dagger},
\end{equation}
where,
\begin{equation}
\begin{aligned}
    M^{\mathrm{amp}}_{j,1}&=\left( \bra{j}^{(1)} V_s \ket{+}_0^{(2)}\right)\otimes \mathds{1} \otimes \mathds{1},\\
      M^{\mathrm{amp}}_{j,2}&=\mathds{1} \otimes\left( \bra{j}^{(1)} V_s \ket{+}_0^{(2)}\right)\otimes  \mathds{1},\\
      M^{\mathrm{amp}}_{j,3}&=\mathds{1} \otimes  \mathds{1} \otimes\left( \bra{j}^{(1)} V_s \ket{+}_0^{(2)}\right), \\
\end{aligned} 
\end{equation}
and  $V_s$ is the unitary operator given in \cref{eq:V_s_gate}.
The Kraus operator representation of the phase error correction for spin-cat encoding is,
\begin{equation}
    \mathcal{R}_{\mathrm{ph}}(\rho)=\sum_{i,j} M_{i,j}^{\mathrm{ph}} \rho \left(M_{i,j}^{\mathrm{ph}}\right)^{\dagger}
\end{equation}
where, 
\begin{equation}
\begin{aligned}
    M_{00}^{\mathrm{ph}}&= \sum_{i,j,k}\ket{+}_i\ket{+}_j\ket{+}_k\bra{+}_i\bra{+}_j\bra{+}_k+\ket{-}_i\ket{-}_j\ket{-}_k\bra{-}_i\bra{-}_j\bra{-}_k,\\
    M_{01}^{\mathrm{ph}}&= Z_3\sum_{i,j,k}\ket{+}_i\ket{+}_j\ket{-}_k\bra{+}_i\bra{+}_j\bra{-}_k+\ket{-}_i\ket{-}_j\ket{+}_k\bra{-}_i\bra{-}_j\bra{+}_k,\\
     M_{10}^{\mathrm{ph}}&= Z_1\sum_{i,j,k}\ket{-}_i\ket{+}_j\ket{+}_k\bra{-}_i\bra{+}_j\bra{+}_k+\ket{+}_i\ket{-}_j\ket{-}_k\bra{+}_i\bra{-}_j\bra{-}_k,\\
      M_{11}^{\mathrm{ph}}&= Z_2\sum_{i,j,k}\ket{+}_i\ket{-}_j\ket{+}_k\bra{+}_i\bra{-}_j\bra{+}_k
      +\ket{-}_i\ket{+}_j\ket{-}_k\bra{-}_i\bra{+}_j\bra{-}_k.\\
\end{aligned}    
\end{equation}
To prove the commutativity of the two error correction steps first consider the Kraus operators $M_{00}^{\mathrm{ph}}$ and $M^{\mathrm{amp}}_{j,1}$, we get
\begin{equation}
    \begin{aligned}       
    M_{j,1}^{\mathrm{amp}}M_{00}^{\mathrm{ph}}=\sum_{klm} \bra{j}\ket{+}_k\left(\ket{+}_0^{(b)}\ket{+}_l^{(a)}\ket{+}_m^{(a)}\bra{+}_k^{(a)}\bra{+}_l^{(a)}\bra{+}_m^{(a)}+\ket{-}_0^{(b)}\ket{-}_l^{(a)}\ket{-}_m^{(a)}\bra{-}_k^{(a)}\bra{-}_l^{(a)}\bra{-}_m^{(a)}\right),\\
    \end{aligned}
\end{equation}
and,
\begin{equation}
\begin{aligned}
     M_{00}^{\mathrm{ph}}M_{j,1}^{\mathrm{amp}}&= \sum_{k,l,m}\ket{+}_k^{(b)}\ket{+}_l^{(a)}\ket{+}_m^{(a)}\bra{+}_k^{(b)}\bra{+}_l^{(a)}\bra{+}_m^{(a)}\bra{j}^{(a)}\otimes \mathds{1}^{(b)} V_s^{(ab)} \mathds{1}^{(a)}\otimes \ket{+}_0^{(b)}\\     &+\sum_{k,l,m}\ket{-}_k^{(b)}\ket{-}_l^{(a)}\ket{-}_m^{(a)}\bra{-}_k^{(b)}\bra{-}_l^{(a)}\bra{-}_m^{(a)}\bra{j}^{(a)}\otimes \mathds{1}^{(b)} V_s^{(ab)} \mathds{1}^{(a)}\otimes \ket{+}_0^{(b)}.\\ 
\end{aligned}   
\end{equation}
Using the resolution of the identity $ \mathds{1}=\sum_p \ket{+}_p\bra{+}_p+\ket{-}_p\bra{-}_p$ in the above equation yields,
\begin{equation}
    \begin{aligned}     M_{00}^{\mathrm{ph}}M_{j,1}^{\mathrm{amp}}=M_{j,1}^{\mathrm{amp}}M_{00}^{\mathrm{ph}}.
\end{aligned} 
\end{equation}
Thus these two Kraus operators commute with each other. 
Similarly, one can find that,
\begin{equation}
    \begin{aligned}
        \comm{M_{j,2}^{\mathrm{amp}}}{M_{00}^{\mathrm{ph}}}&=0,
         \comm{M_{j,3}^{\mathrm{amp}}}{M_{00}^{\mathrm{ph}}}&=0.\\
    \end{aligned}
\end{equation}
Similar calculations also give,
\begin{equation}
    \begin{aligned}
         \comm{M_{j,2}^{\mathrm{amp}}}{M_{10}^{\mathrm{ph}}}&=0,
         \comm{M_{j,2}^{\mathrm{amp}}}{M_{01}^{\mathrm{ph}}}&=0,\\
         \comm{M_{j,3}^{\mathrm{amp}}}{M_{01}^{\mathrm{ph}}}&=0,
         \comm{M_{j,3}^{\mathrm{amp}}}{M_{11}^{\mathrm{ph}}}&=0,\\
         \comm{M_{j,1}^{\mathrm{amp}}}{M_{10}^{\mathrm{ph}}}&=0,
         \comm{M_{j,1}^{\mathrm{amp}}}{M_{10}^{\mathrm{ph}}}&=0.\\
    \end{aligned}
\end{equation}

\noindent Next, consider the Kraus operators, $M_{10}^{\mathrm{ph}}$ and $M^{\mathrm{amp}}_{j,1}$ we get,
\begin{equation}
    \begin{aligned}      
    M_{j,1}^{\mathrm{amp}}M_{10}^{\mathrm{ph}}&=\sum_{klm} (-1)^k\bra{j}\ket{+}_k\left(\ket{+}_0^{(b)}\ket{+}_l^{(a)}\ket{+}_m^{(a)}\bra{-}_k^{(a)}\bra{+}_l^{(a)}\bra{+}_m^{(a)}+\ket{-}_0^{(b)}\ket{-}_l^{(a)}\ket{-}_m^{(a)}\bra{+}_k^{(a)}\bra{-}_l^{(a)}\bra{-}_m^{(a)}\right),\\    M_{10}^{\mathrm{ph}}M_{j,1}^{\mathrm{amp}}&=\sum_{k,l,m}\bra{j}\ket{+}_k\left(\ket{-}_0^{(b)}\ket{+}_l^{(a)}\ket{+}_m^{(a)}\bra{-}_k^{(a)}\bra{+}_l^{(a)}\bra{+}_m^{(a)}+\ket{-}_0^{(b)}\ket{-}_l^{(a)}\ket{-}_m^{(a)}\bra{+}_k^{(a)} \bra{-}_l^{(a)}\bra{-}_m^{(a)}\right),\\
     &\neq M_{j,1}^{\mathrm{amp}}M_{10}^{\mathrm{ph}}.
\end{aligned} 
\end{equation}
Thus these two Kraus operators do not commute with each other, however looking at the full recovery operation, 
\begin{equation}
    \begin{aligned}
       \sum_{j}  M_{10}^{\mathrm{ph}}M_{j,1}^{\mathrm{amp}}  \rho \left(M_{j,1}^{\mathrm{amp}} \right)^{\dagger} \left(M_{10}^{\mathrm{ph}}\right)^{\dagger}
       &=\sum_j \sum_{k,l,m,k',l',m'} \bra{+}_{k'}\ket{j}\bra{j}\ket{+}_kA_{k,l,m} \rho A_{k',l',m'}^{\dagger},\\
       &=\sum_{k,l,m,l',m'} A_{k,l,m} \rho A_{k,l',m'}^{\dagger},
    \end{aligned}
    \label{eq:non_comm_1}
\end{equation}
where we have defined,
\begin{equation}    
\begin{aligned}    A_{k,l,m}&\equiv\ket{-}_0^{(b)}\ket{+}_l^{(a)}\ket{+}_m^{(a)}\bra{-}_k^{(a)}\bra{+}_l^{(a)}\bra{+}_m^{(a)}
&+\ket{-}_0^{(b)}\ket{-}_l^{(a)}\ket{-}_m^{(a)}\bra{+}_k^{(a)} \bra{-}_l^{(a)}\bra{-}_m^{(a)}.
\end{aligned}
\end{equation}
Similarly, we get,
\begin{equation}
\begin{aligned}
       \sum_{j}  M_{j,1}^{\mathrm{amp}}M_{10}^{\mathrm{ph}}  \rho  \left(M_{10}^{\mathrm{ph}}\right)^{\dagger}\left(M_{j,1}^{\mathrm{amp}} \right)^{\dagger}
    &=\sum_j \sum_{k,l,m,k',l',m'} (-1)^{k+k'}\bra{+}_{k'}\ket{j}\bra{j}\ket{+}_kA_{k,l,m} \rho A_{k',l',m'}^{\dagger},\\
    &=\sum_{k,l,m,l',m'} A_{k,l,m} \rho A_{k,l',m'}^{\dagger}.
\end{aligned}
\label{eq:non_comm_2}
\end{equation}
Combining \cref{eq:non_comm_1} and \cref{eq:non_comm_2} gives,
\begin{equation}
\begin{aligned}
      \sum_{j} M_{j,1}^{\mathrm{amp}}M_{10}^{\mathrm{ph}}  \rho  \left(M_{10}^{\mathrm{ph}}\right)^{\dagger}\left(M_{j,1}^{\mathrm{amp}} \right)^{\dagger}
      =\sum_{j}M_{10}^{\mathrm{ph}}M_{j,1}^{\mathrm{amp}}  \rho \left(M_{j,1}^{\mathrm{amp}} \right)^{\dagger} \left(M_{10}^{\mathrm{ph}}\right)^{\dagger}
\end{aligned}  
\end{equation}
Similarly one can find,
\begin{equation}
    \begin{aligned}
        \sum_{j}M_{j,2}^{\mathrm{amp}}M_{11}^{\mathrm{ph}}  \rho  \left(M_{11}^{\mathrm{ph}}\right)^{\dagger}\left(M_{j,2}^{\mathrm{amp}} \right)^{\dagger}
        &=\sum_{j}M_{11}^{\mathrm{ph}}M_{j,2}^{\mathrm{amp}}  \rho \left(M_{j,1}^{\mathrm{amp}} \right)^{\dagger} \left(M_{01}^{\mathrm{ph}}\right)^{\dagger}\\
       \sum_{j} M_{j,3}^{\mathrm{amp}}M_{10}^{\mathrm{ph}}  \rho  \left(M_{01}^{\mathrm{ph}}\right)^{\dagger}\left(M_{j,3}^{\mathrm{amp}} \right)^{\dagger}
       &=\sum_{j}M_{01}^{\mathrm{ph}}M_{j,3}^{\mathrm{amp}}  \rho \left(M_{j,3}^{\mathrm{amp}} \right)^{\dagger} \left(M_{01}^{\mathrm{ph}}\right)^{\dagger}
    \end{aligned}
\end{equation}     
Combining all these we get,
\begin{equation}
     \mathcal{R}_{\mathrm{amp}} \mathcal{R}_{\mathrm{ph}}\left(\mathcal{E}\left(\rho\right)\right)= \mathcal{R}_{\mathrm{ph}} \mathcal{R}_{\mathrm{amp}}\left(\mathcal{E}\left(\rho\right)\right),
\end{equation}
and thus the phase error correction and amplitude error correction commute with each other.
\end{widetext}

\bibliography{reference}
\end{document}